\documentclass{lmcs}
\usepackage[utf8]{inputenc}
\usepackage{textcomp}
\usepackage{amssymb,bm,graphicx,array}
\usepackage{stmaryrd}
\SetSymbolFont{stmry}{bold}{U}{stmry}{m}{n}
\usepackage{hyperref}
\usetikzlibrary{arrows.meta,positioning,decorations.pathreplacing}

\newcommand{\Nat}{\mathbb{N}}

\newcommand{\Rat}{\mathbb{Q}}
\newcommand{\Real}{\mathbb{R}}

\newcommand{\Ralg}{\mathbb{R}_{\mathrm{alg}}}
\newcommand{\EL}{\mathsf{EL}}
\newcommand{\ELR}{\mathsf{EL}(\mathbb{R}^k)}

\newcommand{\RCF}{\mathsf{RCF}}
\newcommand{\Prop}{\mathsf{P}}
\newcommand{\GE}{\mathsf{GE}}
\newcommand{\boldBox}{\Box}
\newcommand{\mdiamond}{\Diamond}
\newcommand{\sem}[1]{\llbracket #1\rrbracket}
\newcommand{\norm}[1]{\lVert #1\rVert}
\newcommand{\inner}[2]{\langle #1,#2\rangle}
\newcommand{\dist}{\operatorname{dist}}
\newcommand{\rad}{\operatorname{rad}}
\newcommand{\Intr}{\mathsf{int}}
\newcommand{\Clo}{\mathsf{cl}}
\newcommand{\Full}{\mathsf{Full}}
\newcommand{\Cohort}{\mathsf{Cohort}}
\newcommand{\Bool}{\mathsf{Bool}}
\newcommand{\tp}{\mathsf{T}}
\definecolor{inkblue}{RGB}{31,78,140}
\definecolor{inkred}{RGB}{160,40,40}
\definecolor{inkgreen}{RGB}{30,110,60}

\newif\ifrasters
\newcommand{\rasterdir}{}
\IfFileExists{fig4_angiography.pdf}{\rasterstrue}{%
  \IfFileExists{images/fig4_angiography.pdf}{\rasterstrue\renewcommand{\rasterdir}{images/}}{%
    \IfFileExists{figures/fig4_angiography.pdf}{\rasterstrue\renewcommand{\rasterdir}{figures/}}{}}}

\hypersetup{pdfsubject={Spatial Ensemble Logic and Geometric Element Ensembles -- Version 8}}

\begin{document}

\title[Spatial Ensemble Logic and Geometric Element Ensembles]{Spatial Ensemble
Logic for Biomedical Imaging: Geometric Element Ensembles in Euclidean Space}
\author[G.-Q.~Zhang]{Guo-Qiang Zhang}
\address{The University of Texas Health Science Center at Houston, Houston, Texas, USA}
\email{guo-qiang.zhang@uth.tmc.edu}
\ACMCCS{Theory of computation $\rightarrow$ Modal and temporal logics;
Logic and verification; Computational geometry}
\date{}
\begin{abstract}
Ensemble Logic (EL) is a family of modal logics over metric monoids whose
modal parameters are terms with quantified variables. We study its Euclidean
member $\ELR$, Spatial Ensemble Logic (SEL), in dimension $k\geq2$, as a
language for the geometric properties of biomedical images. Interpretations
are presented by geometric element ensembles (GELs), which assign to each
atomic proposition a finite list of effectively presented semi-algebraic
sets. Closed balls are the principal elements. The passage from the
interval ensembles of temporal annotation ($k=1$) to balls ($k\geq 2$) encounters a problem of
closure. Translation, dilation and erosion carry a ball to a ball or the empty set
in any dimension. Translation and dilation act on an ensemble element by element.
Element-wise erosion gives only an inner bound; for $k\geq 2$ finite unions
of balls are closed neither under intersection nor under erosion. We therefore separate the
representation of a record from the domain in which formulas are evaluated.
The main result is a translation of $\ELR$ over GELs into the first-order
theory of real closed fields that preserves satisfaction. It yields
computable semi-algebraic satisfaction sets, decidable model checking and
finite real algebraic evidence. For a fixed formula and dimension, model
checking has polynomial data complexity under stated assumptions of
rationality and bounded degree. The combined complexity is PSPACE-hard and in
EXPSPACE. On the line, rational interval data admit evaluation in linear
arithmetic with rational evidence, under rationality assumptions on the
norm, the query and the point of evaluation. For $k\geq2$ rational balls can
force irrational witnesses. Lower and upper bounds on an annotation
propagate through formulas. Synthesis of a ball ensemble is decidable within
a prescribed budget of balls and undecidable without one.
In a vascular use case, medial ball ensembles decide calibre and stenosis
queries at stored centres by comparing stored numbers. We also examine how
far evaluation can proceed on the elements when the union that forms a
support is deferred to the point of evaluation.
\end{abstract}
\maketitle

\section{Introduction}\label{sec:intro}

A segmentation of a biomedical image assigns anatomical or tissue labels to
locations. Questions about it usually concern their spatial relations:
whether a lesion is near a vessel, whether a region free of vessels contains
a ball of a given size, whether a structure persists throughout a
neighbourhood. Answers depend on choices that an image-processing pipeline
tends to leave implicit. These concern the physical distance between samples,
boundaries and the field of view, and whether several conditions are to hold
at one location or merely close together. A logical specification makes
these choices part of the object that is evaluated.

\emph{Ensemble Logic} ($\EL$) is a framework for such
specifications~\cite{zhang2024temporal}. Its formulas are interpreted
over a metric monoid, whose elements are called points. Displacement
$\varphi_u$ evaluates $\varphi$ at the point reached by adding the value of
$u$ to the current point. Bounded existence $\mdiamond_t\varphi$ and bounded
universality $\boldBox_t\varphi$ evaluate $\varphi$ at some, respectively at
every, point of a window whose extent is the value of $t$. The parameters
$u$ and $t$ are terms, and their variables may be quantified over points and
over lengths. For rational time this design underlies a method of cohort
discovery, Rational Ensemble Logic (QEL)~\cite{qel}. A labelled interval
ensemble (LINE) represents an annotated recording. A formula describes a
temporal pattern, and a model checker decides whether the recording exhibits
it. The general theory~\cite{zhang2024temporal} treats the temporal members of the family
in depth and leaves open which spatial members admit effective model
checking on finite geometric presentations such as unions of balls.

We answer that question for Euclidean space in arbitrary dimension. A
\emph{geometric element ensemble} (GEL) presents an interpretation by finitely
many simple geometric elements. It plays the role for an annotated image
that a LINE plays for an annotated recording. We connect Ensemble Logic over
GELs to the first-order theory of real closed fields. This gives the spatial
logic, with its quantifiers of both sorts, a computable semantics, an exact
decision procedure and evidence for its verdicts.

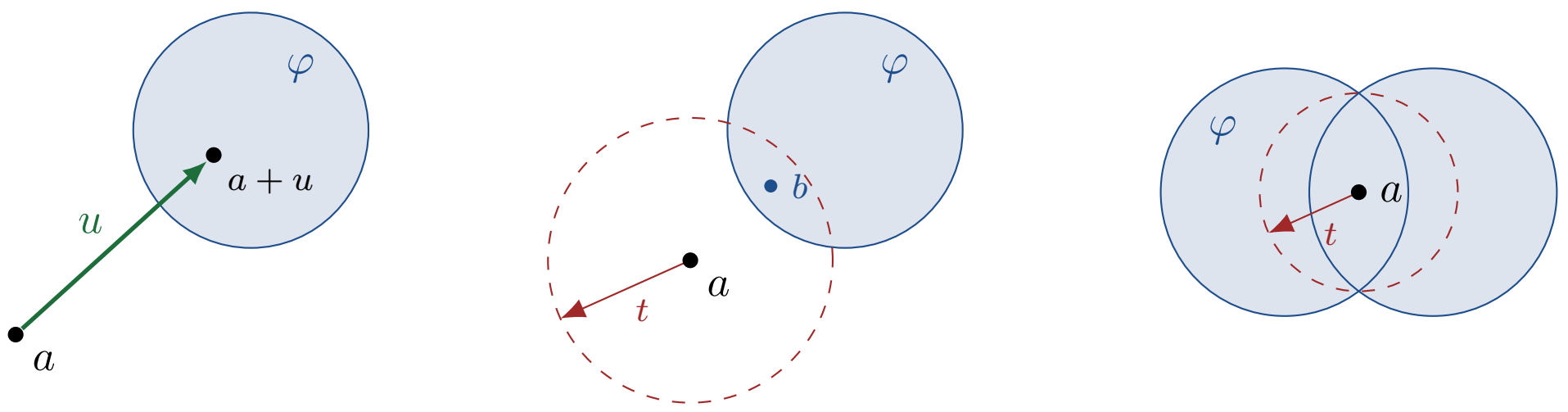
\begin{figure}[htbp]
\centering
\begin{tikzpicture}[font=\small,>={Latex[length=2mm,width=1.6mm]}]
\begin{scope}
\fill[inkblue!15] (2.45,2.05) circle (0.95);
\draw[inkblue] (2.45,2.05) circle (0.95);
\node[inkblue] at (2.85,2.55){$\varphi$};
\draw[->,inkgreen,line width=1pt,shorten >=2pt,shorten <=2pt] (0.55,0.4)--(2.15,1.85)
  node[midway,above left=-2pt]{$u$};
\fill (0.55,0.4) circle (1.8pt) node[below right=0pt]{$a$};
\fill (2.15,1.85) circle (1.8pt) node[below right=-1pt]{\scriptsize $a+u$};
\node at (1.9,-0.55){(a) $\varphi_u$};
\end{scope}
\begin{scope}[shift={(4.7,0)}]
\fill[inkblue!15] (2.55,2.05) circle (0.95);
\draw[inkblue] (2.55,2.05) circle (0.95);
\node[inkblue] at (2.95,2.55){$\varphi$};
\draw[dashed,inkred] (1.3,1.0) circle (1.15);
\draw[->,inkred] (1.3,1.0)--(0.25,0.53) node[midway,below right=-2pt]{\scriptsize $t$};
\fill (1.3,1.0) circle (1.8pt) node[below right=0pt]{$a$};
\fill[inkblue] (1.95,1.6) circle (1.5pt) node[right=1pt]{\scriptsize $b$};
\node at (1.9,-0.55){(b) $\mdiamond_t\varphi$};
\end{scope}
\begin{scope}[shift={(9.6,0)}]
\fill[inkblue!15] (1.2,1.55) circle (1.0) (2.4,1.55) circle (1.0);
\draw[inkblue] (1.2,1.55) circle (1.0) (2.4,1.55) circle (1.0);
\node[inkblue] at (0.7,2.05){$\varphi$};
\draw[dashed,inkred] (1.8,1.55) circle (0.8);
\draw[->,inkred] (1.8,1.55)--(1.07,1.22) node[midway,below right=-2pt]{\scriptsize $t$};
\fill (1.8,1.55) circle (1.8pt) node[right=1pt]{$a$};
\node at (1.8,-0.55){(c) $\boldBox_t\varphi$};
\end{scope}
\end{tikzpicture}
\caption{The three modal operators of $\ELR$ at a point $a$, drawn in the
plane. A shaded region is the support of $\varphi$, and a dashed circle
bounds a window, which is an open ball. (a)~Displacement. (b)~Bounded
existence, with a witness $b$. (c)~Bounded universality; here two closed
balls cover the window jointly, and neither contains it.}
\label{fig:operators}
\end{figure}

The space is $\Real^k$ for a positive integer $k$, with a metric expressing
physical distance in image coordinates. A profile, a planar image and a
volume are the cases $k=1$, $2$ and $3$. A \emph{record} is a GEL: to an
atomic proposition of a finite vocabulary, such as lesion, vessel or
observed region, it assigns a finite list of finitely described geometric
elements. Closed balls are the principal kind. The atomic proposition holds
on their union. A
\emph{specification} is a formula of Spatial Ensemble Logic (SEL), the
member $\ELR$ of the family $\EL$ in which the window of length $t$ at a
point $a$ is the open ball of radius $t$ about $a$
(Figure~\ref{fig:operators}). The \emph{support} of a formula is the set of
points at which it holds. \emph{Evaluation}, or model checking, decides
whether a given point of a record belongs to it. Applying one formula to the
records of a database selects a cohort.
Sections~\ref{sec:framework} and~\ref{sec:gel} give the definitions in full.

If $L$ and $V$ label a lesion and a vessel, then $L\wedge\mdiamond_sV$ holds
at lesion points that have a vessel point at a distance less than $s$.
The formula $\boldBox_rV$ holds where a ball of radius $r$ fits inside the
vessel region. Quantifying a displacement turns the first, local condition
into a statement about the record: $\exists x\,(L\wedge\mdiamond_sV)_x$ says
that some lesion point has a vessel point less than $s$ away. Quantified
lengths let different windows share a scale: with one unit $z$, the formulas
$\neg\boldBox_{3z}V$ at a stenosis and $\boldBox_{10z}V$ at a distal
reference site compare two calibres, where $nz$ abbreviates an $n$-fold sum
(Section~\ref{sec:vascular}).

The main obstacle in passing from time to space is closure. QEL evaluates
formulas on finite unions of rational intervals, a class closed under the
operations of the logic. The support of a subformula is therefore again a
record. Balls retain part of this advantage: translation, dilation and
erosion carry a ball to a ball or to the empty set in any dimension. Finite
unions of balls need not retain it. For $k\geq2$ the intersection of two balls is a lens, and a
window may be covered by several balls jointly while none of them contains
it, as in Figure~\ref{fig:operators}(c). A calculus confined to finite
unions of balls therefore cannot represent the supports of all subformulas.

We keep balls as the elements of a record and evaluate formulas in the
larger class of semi-algebraic sets. These are defined by finite Boolean
combinations of polynomial equalities and inequalities. They are closed
under Boolean operations and quantification by the Tarski--Seidenberg
theorem~\cite{tarski1951,seidenberg1954,bpr2006}. Their closure under
translation, dilation and erosion by open balls was observed
in~\cite{qel}. We develop that observation into model checking for the
full spatial logic. The main theorem connects the representation of a record
with the domain of evaluation: a formula of $\ELR$ over a GEL translates
into the first-order theory of real closed fields with satisfaction
preserved
(Theorem~\ref{thm:translation}). Consequently the satisfying points and
parameter values form a computable semi-algebraic set, model checking is
decidable, and each block of like quantifiers has finite real algebraic
evidence (Theorem~\ref{thm:evidence}). For a fixed formula and dimension,
with rational calibration and a fixed degree bound on rational input, model
checking takes time polynomial in the binary size of the input
(Theorem~\ref{thm:datacomplexity}). A spatial phenotype thus has a
semantics, a decision procedure and checkable evidence that do not depend on
the data structure in which the annotation is stored.

The line shows where simpler methods suffice. For $k=1$ balls are
intervals, and their finite unions are closed under intersection and erosion.
Rational interval input is evaluated in linear arithmetic, with rational
evidence under the rationality assumptions of Proposition~\ref{prop:linear},
as in QEL. For $k\geq2$ none of these conclusions holds in general. The
definitions are the same in all dimensions; the geometry decides which
simplifications survive.

\subsection{Contributions}\label{subsec:contrib}
The results fall into five areas. The first two prepare the model-checking
result in the third; the last two develop its uses and delimit its scope.

\begin{enumerate}[leftmargin=2em]
\item \emph{The spatial logic and its finite presentations}
(Sections~\ref{sec:framework} and~\ref{sec:gel}). The logic $\ELR$ is the
member of the family $\EL$ of~\cite{zhang2024temporal} for the metric monoid $\Real^k$ with
a calibrated Euclidean metric, in full mode. It inherits the laws and the
atomic-displacement normal form of the general theory
(Proposition~\ref{prop:laws}). A GEL presents an interpretation by a finite
list of effective semi-algebraic elements for each atomic proposition
(Definition~\ref{def:gel}).
\item \emph{Spatial morphology and the role of the dimension}
(Section~\ref{sec:morphology}). The modal operators act on supports as
translation, dilation and erosion (Propositions~\ref{prop:semantics}
and~\ref{prop:morph}). Balls are closed under these operations
(Proposition~\ref{prop:ball}), which act on a ball ensemble element by
element, exactly for translation and dilation and as an inner bound for
erosion (Proposition~\ref{prop:elementwise}). Finite unions of balls are
closed under intersection and erosion on the line and under neither for
$k\geq2$ (Proposition~\ref{prop:closure}), and a window is a finite union
of smaller windows only on the line (Proposition~\ref{prop:windows}).
Derived topological operators (Proposition~\ref{prop:topology},
Corollary~\ref{cor:connected}) and cylindrical transfer between dimensions
(Theorem~\ref{thm:cylinder}) complete the account.
\item \emph{Exact model checking} (Section~\ref{sec:modelchecking}). The
standard translation into real closed fields
(Theorem~\ref{thm:translation}) yields semi-algebraic satisfaction sets and
finite evidence (Theorem~\ref{thm:evidence}), polynomial data complexity
under the stated rationality and degree assumptions
(Theorem~\ref{thm:datacomplexity}), and bounds on the combined complexity
(Proposition~\ref{prop:combined}). For rational polyhedral norms and
rational semilinear input, including rational interval data on the line,
evaluation remains in linear arithmetic (Proposition~\ref{prop:linear}),
whereas Euclidean balls in dimension $k\geq2$ produce irrational witnesses
from rational data (Example~\ref{ex:irrational-length}).
\item \emph{Approximate records and synthesis}
(Sections~\ref{sec:enclosures} and~\ref{sec:synthesis}). Lower and upper
bounds on atomic supports propagate soundly through formulas
(Theorem~\ref{thm:enclosure}). We also characterize the bounds obtainable
from a tolerance (Proposition~\ref{prop:tolerance}). Whether some ball-GEL
within a prescribed budget of balls satisfies a given formula is decidable
(Theorem~\ref{thm:synthesis}); without the budget, satisfiability over GELs
is $\Sigma^0_1$-complete, and satisfiability over arbitrary interpretations
is $\Pi^0_1$-hard (Proposition~\ref{prop:limits}).
\item \emph{A vascular use case, and evaluation on the elements}
(Sections~\ref{sec:vascular}--\ref{sec:discussion}). Medial ball ensembles
give exact calibre verdicts at stored centres (Lemma~\ref{lem:medial}) and
obey a sampling law along chains (Lemma~\ref{lem:chain}); a library of
phenotypes includes stenosis grading by a ratio of calibres, and exact
synthetic diagnostics fix the boundary conventions. The Discussion shows how
far a verdict can be computed on the elements when the union that forms a
support is taken last.
\end{enumerate}

\noindent
The claims of the paper are mathematical and algorithmic. The vascular use
case supplies geometric specifications and an evaluation protocol. It
reports no clinical findings, experiments on patient data or validated
diagnostic thresholds.

\subsection{Related work}\label{subsec:related}
\emph{Ensemble Logic.} The operators of $\EL$ were introduced for linear
time~\cite{zhang2024temporal}; the discrete logic over the natural numbers
is studied in joint work with Droste~\cite{DZ26}, and~\cite{zhang2024temporal} develops the
family over metric monoids. QEL~\cite{qel} is our closest predecessor in
method. We keep its separation of model, formula and checker, while full
open balls replace the future windows $[a,a+\ell)$, ensembles of
semi-algebraic elements replace rational intervals, and real closed fields
take the place of ordered divisible abelian groups in exact evaluation.

\emph{Mathematical morphology and spatial logics.} Dilation and erosion by
a structuring element are the basic operations of mathematical
morphology~\cite{matheron1975,serra1982,heijmans1994}, discussed further in
Section~\ref{subsec:contrast}. Modal logics based on these operations are
studied in~\cite{bloch2002,bhr2007}. Kutz et al.~\cite{kutz2003} studied modal
operators that quantify over the points within a prescribed distance in
logics of metric spaces. The language $\EL$ adds displacement by affine
terms, lengths given by terms, and quantification over both sorts. Region-based
formalisms such as the Region Connection Calculus~\cite{cohn} take regions
as primitive~\cite{aiello2007}. Model checking over closure spaces, which
include digital images~\cite{cllm2016}, is applied to medical-image analysis
in VoxLogicA~\cite{belmonte2019}. A voxel model is a finite computational
domain, whereas a GEL describes a region between the samples
(Section~\ref{subsec:cells}).

\emph{Constraint databases.} The evaluation of first-order queries over
finitely presented semi-algebraic relations by quantifier elimination is a
central construction of constraint databases~\cite{kkr1995}. In these terms
a GEL is a constraint database with one $k$-ary relation for each atomic
proposition, and our translation compiles $\ELR$ into a first-order query.
Specific to the present setting are the correctness of this compilation for
a modal language with variable parameters, finite evidence, the element-wise
calculus of balls with its limits, and sound evaluation on approximate
records.

\subsection{Organization}
The sections follow the order of the contributions, ending with
Section~\ref{sec:conclusion}. Proofs accompany their statements because
later sections reuse their constructions. Statements marked ``cf.''\ are
instances of the general theory.

\section{Ensemble Logic over Euclidean Space}\label{sec:framework}

The spatial theory retains the logical primitives of~\cite{zhang2024temporal}. Its
specific features are the metric monoid $\Real^k$ and the windows through
which formulas inspect it. The definitions specialize those
of~\cite{zhang2024temporal} uniformly in $k$. The laws of
Section~\ref{subsec:laws} support the calculations that follow.

\subsection{The metric monoid \texorpdfstring{$\Real^k$}{Rk}}\label{subsec:mm}
Throughout the paper $k$ is a positive integer, the \emph{dimension}. We
write $a=(a_1,\ldots,a_k)$ for the elements of $\Real^k$, $e_1,\ldots,e_k$
for the standard unit vectors, $0$ for the origin, and $\Real_{>0}$ for the
set of positive reals. A \emph{metric monoid}, or a monoid~\cite{zhang2024temporal} is a commutative
monoid $(M,+,0)$ with a metric $d$ that is translation invariant,
$d(a+c,b+c)=d(a,b)$, considered together with a \emph{length domain}: a
nonempty set of positive reals that is closed under addition and contains
the distance $d(a,0)$ of every element $a\neq0$.

\begin{defi}[Euclidean space]\label{def:space}
Let $G$ be a symmetric positive-definite real $k\times k$ matrix, the
\emph{metric matrix}. For $v,w\in\Real^k$ put
\[
\inner{v}{w}:=v^{\tp}Gw,\qquad
\norm{v}:=\inner{v}{v}^{1/2},\qquad
d(a,b):=\norm{a-b}.
\]
\emph{Euclidean space} of dimension $k$ is the metric monoid
$(\Real^k,+,0,d)$ with the length domain $\Real_{>0}$; we write $\Real^k$ for
it. Its elements are called \emph{points}, and the positive reals are called
\emph{lengths}. The \emph{norm} of a point $a$ is $\norm{a}=d(a,0)$. For a
point $c$, the \emph{open ball} of radius $\ell>0$ and the \emph{closed ball}
of radius $\rho\geq0$ about $c$ are
\[
B(c,\ell):=\{b\in\Real^k\mid\norm{b-c}<\ell\}
\qquad\text{and}\qquad
D(c,\rho):=\{b\in\Real^k\mid\norm{b-c}\leq\rho\}.
\]
\end{defi}

\noindent
Points serve as the locations at which formulas are evaluated and as the
displacements between them. A closed ball may have the radius $0$, and the
singleton $D(c,0)=\{c\}$ is needed for exact assertions about points. For
the identity matrix $G=I$ the norm is the standard one, written
$|v|:=(v_1^2+\cdots+v_k^2)^{1/2}$. In dimension one, $G=(g)$ with $g>0$ and
$\norm{v}=\gamma|v|$ for $\gamma:=g^{1/2}$, so that balls are bounded
intervals. Two vectors $w_1,w_2$ form an \emph{orthonormal pair} if
$\norm{w_1}=\norm{w_2}=1$ and $\inner{w_1}{w_2}=0$. Such a pair exists
exactly when $k\geq2$. The results below use the hypothesis $k\geq2$ only
to obtain such a pair; for $k=1$ they use the fact that a vector of norm $1$
spans the space.

The metric matrix carries the calibration of an image into the semantics.
In Cartesian physical coordinates $G=I$; if the stored image frame is mapped
to physical coordinates by a linear map $M$, then $G=M^{\tp}M$, and
axis-aligned sample spacings $\sigma_1,\ldots,\sigma_k$ give the diagonal
matrix with the entries $\sigma_i^2$. Different metric matrices describe
isometric spaces. When $G=M^{\tp}M$, the map $v\mapsto Mv$ carries the norm
of $G$ to the standard norm. The space $\Real^k$ with the standard norm is
the Euclidean metric monoid listed in~\cite{zhang2024temporal}.
Proposition~\ref{prop:coordinates} extends the isometry to the logic. If
$\lambda_{\min}$ and $\lambda_{\max}$ are the extreme eigenvalues of $G$,
then $\lambda_{\min}|v|^2\leq\norm{v}^2\leq\lambda_{\max}|v|^2$; hence $d$
induces the usual topology of $\Real^k$, in which interiors, closures and
boundaries are taken.

For sets $S,T\subseteq\Real^k$ and a point $c$ we write
$S^{c}:=\Real^k\setminus S$ for the complement,
$S+T:=\{a+b\mid a\in S,\ b\in T\}$ for the Minkowski sum,
$S+c:=\{a+c\mid a\in S\}$ and $S-c:=\{a-c\mid a\in S\}$ for the translates,
and $\dist(a,S):=\inf\{d(a,b)\mid b\in S\}$ for the distance of a point $a$
from $S$, with $\dist(a,\emptyset)=\infty$. We write $\Ralg$ for the field of
real algebraic numbers, each given effectively by a defining polynomial with
integer coefficients and an isolating interval with rational
endpoints~\cite{bpr2006}. The effective results below assume that the
entries of $G$ lie in $\Ralg$, and the complexity bounds assume a rational
$G$.

For $k=1$ and $G=(1)$ we recover the signed linear metric monoid $\Real$
of~\cite{zhang2024temporal}, though with the spatial window: the interval $(a-\ell,a+\ell)$
is symmetric about $a$, whereas the linear-time logics of~\cite{qel} use
the future window $[a,a+\ell)$. The field of view of an image is
represented by an atomic proposition (Section~\ref{subsec:observation}),
since a bounded rectangle is not closed under addition and cannot serve as
the metric monoid.

\subsection{Syntax}\label{subsec:syntax}
Terms come in two sorts, for the displacement of a location and for the
extent of a window. Following the general theory, we call terms of the point
sort affine and terms of the length sort positive affine.

\begin{defi}[Affine terms]\label{def:terms}
Fix disjoint countable sets of \emph{point variables} $x,y$ and of
\emph{length variables} $z,w$. \emph{Affine terms} $u,v$ and \emph{positive
affine terms} $s,t$ are given by
\[
u,v\;::=\;c\mid x\mid u+v
\qquad\text{and}\qquad
s,t\;::=\;r\mid z\mid s+t\mid s+\norm{u},
\]
where $c$ ranges over constants for the points of $\Ralg^k$ and $r$ over
constants for the lengths in $\Ralg$. Each constant names a point or a
length, and we use the same symbol for a constant and for what it names. An
\emph{assignment} $\eta$ maps every point variable to a point and every
length variable to a length. It extends to terms by $\eta(c)=c$,
$\eta(u+v)=\eta(u)+\eta(v)$, $\eta(r)=r$, $\eta(s+t)=\eta(s)+\eta(t)$ and
$\eta(s+\norm{u})=\eta(s)+\norm{\eta(u)}$. Under $\eta$, an affine term $u$
\emph{denotes} the point $\eta(u)$, and a positive affine term $t$ denotes
the length $\eta(t)$. A term is \emph{closed} if it contains no variable.
\end{defi}

\noindent
The grammar ensures that a positive affine term denotes a positive length.
A bare norm $\norm{u}$ is not a term: it would denote $0$ when $u$ denotes
the origin, whereas a window contains its point of evaluation~\cite{zhang2024temporal}. For a positive integer $n$, the
expressions $nu$ and $nt$ denote $n$-fold sums. The syntax has no
subtraction symbol; constants are signed, so that $-c$ is a constant
whenever $c$ is, but the difference $x-y$ of two variables is not an affine
term. In the linear metric monoids of~\cite{zhang2024temporal} a length is also a point.
In dimension $k\geq2$ a length determines no direction of displacement.
We keep the two sorts apart, including for $k=1$, so that the syntax is
uniform across dimensions.

\begin{defi}[Formulas]\label{def:formulas}
Let $\Prop$ be a nonempty set of \emph{atomic propositions}. The
\emph{formulas} of $\EL$ over $\Real^k$ are given by
\[
\varphi,\psi\;::=\;p\mid\neg\varphi\mid\varphi\wedge\psi\mid\varphi\vee\psi\mid
\varphi_u\mid\mdiamond_t\varphi\mid\boldBox_t\varphi\mid
\exists x\,\varphi\mid\forall x\,\varphi\mid\exists z\,\varphi\mid\forall z\,\varphi,
\]
where $p\in\Prop$, $u$ is an affine term, $t$ is a positive affine term, $x$
is a point variable and $z$ is a length variable. The three \emph{modal
operators} are \emph{displacement} $\varphi_u$, \emph{bounded existence}
$\mdiamond_t\varphi$ and \emph{bounded universality} $\boldBox_t\varphi$; the
last two are the \emph{bounded modalities}. The term in the subscript of a
modal operator is its \emph{parameter}. We use the abbreviations
$\top:=p\vee\neg p$ for a fixed $p\in\Prop$, $\bot:=\neg\top$ and
$\varphi\rightarrow\psi:=\neg\varphi\vee\psi$.
\end{defi}

\noindent
Variables occur only in modal parameters, and quantifiers bind their
occurrences within scope. Free and bound occurrences, \emph{closed}
formulas and \emph{quantifier-free} formulas have their usual meanings. A
variable is \emph{fresh} for a formula or term if it does not occur there;
the \emph{size} $|\varphi|$ counts the nodes of the syntax tree, including
term nodes. The atomic propositions are the only predicates: neither $u=v$
nor $s<t$ is a formula.

\subsection{Semantics}\label{subsec:semantics}
An \emph{interpretation} is a map $\alpha\colon\Real^k\to2^{\Prop}$; it
specifies which atomic propositions hold at each point.

\begin{defi}[Satisfaction]\label{def:sat}
Let $\alpha$ be an interpretation, $\eta$ an assignment, and $a$ a point, the
\emph{point of evaluation}. The relation $(\alpha,\eta,a)\models\varphi$,
read \emph{$\varphi$ holds at $a$ under $\alpha$ and $\eta$}, is defined by
the following clauses:
\[
\begin{array}{lcl}
(\alpha,\eta,a)\models p &\text{iff}& p\in\alpha(a);\\
(\alpha,\eta,a)\models \neg\varphi &\text{iff}& (\alpha,\eta,a)\not\models\varphi;\\
(\alpha,\eta,a)\models \varphi\wedge\psi &\text{iff}&
   (\alpha,\eta,a)\models\varphi \text{ and } (\alpha,\eta,a)\models\psi;\\
(\alpha,\eta,a)\models \varphi\vee\psi &\text{iff}&
   (\alpha,\eta,a)\models\varphi \text{ or } (\alpha,\eta,a)\models\psi;\\
(\alpha,\eta,a)\models \varphi_u &\text{iff}& (\alpha,\eta,\,a+\eta(u))\models\varphi;\\
(\alpha,\eta,a)\models \mdiamond_t\varphi &\text{iff}&
   (\alpha,\eta,b)\models\varphi \text{ for some point } b\in B(a,\eta(t));\\
(\alpha,\eta,a)\models \boldBox_t\varphi &\text{iff}&
   (\alpha,\eta,b)\models\varphi \text{ for every point } b\in B(a,\eta(t));\\
(\alpha,\eta,a)\models \exists x\,\varphi &\text{iff}&
   (\alpha,\eta[x\mapsto b],a)\models\varphi \text{ for some point } b;\\
(\alpha,\eta,a)\models \exists z\,\varphi &\text{iff}&
   (\alpha,\eta[z\mapsto\ell],a)\models\varphi \text{ for some length } \ell;
\end{array}
\]
and dually for $\forall x$ and $\forall z$. Here $\eta[\xi\mapsto e]$ agrees
with $\eta$ except that it maps $\xi$ to $e$. The open ball $B(a,\ell)$ is
the \emph{window of length $\ell$ at $a$}.
\end{defi}

\noindent
These are the satisfaction clauses of~\cite{zhang2024temporal} for the \emph{full mode}, in
which the window of length $\ell$ at a point is the whole open ball of radius
$\ell$ about it.

\begin{defi}[Support, equivalence, the logic]\label{def:entail}
The \emph{support} of a formula $\varphi$ under $\alpha$ and $\eta$ is the
set of points at which it holds,
\[
\sem{\varphi}_{\alpha,\eta}:=\{a\in\Real^k\mid(\alpha,\eta,a)\models\varphi\}.
\]
A formula $\varphi$ \emph{entails} $\psi$, written $\varphi\models\psi$, if
$\sem{\varphi}_{\alpha,\eta}\subseteq\sem{\psi}_{\alpha,\eta}$ for all
$\alpha$ and $\eta$; the two are \emph{equivalent}, written
$\varphi\equiv\psi$, if each entails the other. A formula is
\emph{satisfiable} if at least one of its supports is nonempty, and
\emph{valid} if each of its supports is $\Real^k$. The formulas over
$\Real^k$ with this satisfaction relation constitute \emph{the logic
$\ELR$}, which we also call \emph{Spatial Ensemble Logic} (SEL); the metric
matrix, the constants and $\Prop$ are understood from the context.
\end{defi}

\noindent
Satisfaction depends only on the values of $\eta$ on the free variables of
$\varphi$~\cite{zhang2024temporal}. For a closed formula we therefore write
$(\alpha,a)\models\varphi$ and $\sem{\varphi}_\alpha$. That the window is
open can be observed in the logic: if $p$ holds only at a point at distance
exactly $\ell$ from $a$, then $\mdiamond_\ell p$ fails at $a$. Modalities
over closed balls decide such cases differently and do not arise from a
window mode of~\cite{zhang2024temporal}.

\subsection{Laws}\label{subsec:laws}
Because $\Real^k$ is a group, the full mode is
\emph{covariant}~\cite{zhang2024temporal}: $B(a+c,\ell)=B(a,\ell)+c$ for all points $a,c$
and lengths $\ell$. This is the hypothesis of the general laws collected
below. A \emph{literal} is an atomic proposition or its negation. A formula
is in \emph{negation normal form} when negation applies only to atomic
propositions; it is in \emph{atomic-displacement normal form} (ADNF) when,
in addition, every displacement applies to a literal~\cite{zhang2024temporal}.

\begin{prop}[General laws; cf.~\cite{zhang2024temporal}]\label{prop:laws}
For all formulas $\varphi,\psi$, affine terms $u,v$ and positive affine terms
$s,t$:
\begin{enumerate}[label=\textup{(\roman*)},leftmargin=2.2em]
\item \textup{(Translation)}\; for a point $c$ let $\alpha^{c}$ be the
interpretation $b\mapsto\alpha(b+c)$; then $(\alpha^{c},\eta,a)\models\varphi$
if and only if $(\alpha,\eta,a+c)\models\varphi$;
\item \textup{(Displacement)}\; $\varphi_0\equiv\varphi$ and
$(\varphi_u)_v\equiv\varphi_{u+v}$; displacement commutes with the Boolean
connectives and with the bounded modalities, as in
$(\neg\varphi)_u\equiv\neg(\varphi_u)$ and
$(\boldBox_t\varphi)_u\equiv\boldBox_t(\varphi_u)$; and
$(Q\xi\,\varphi)_u\equiv Q\xi\,(\varphi_u)$ for $Q\in\{\exists,\forall\}$ and
every variable $\xi$ that is fresh for $u$;
\item \textup{(Bounded modalities)}\;
$\neg\boldBox_t\varphi\equiv\mdiamond_t\neg\varphi$;
$\boldBox_s\varphi\models\varphi\models\mdiamond_s\varphi$;
$\mdiamond_s\varphi\models\mdiamond_{s+t}\varphi$ and
$\boldBox_{s+t}\varphi\models\boldBox_s\varphi$;
$\mdiamond_t(\varphi\vee\psi)\equiv\mdiamond_t\varphi\vee\mdiamond_t\psi$ and
$\boldBox_t(\varphi\wedge\psi)\equiv\boldBox_t\varphi\wedge\boldBox_t\psi$;
\item \textup{(Reach)}\; $\varphi_u\models\mdiamond_{t+\norm{u}}\varphi$ and
$\boldBox_{t+\norm{u}}\varphi\models\varphi_u$;
\item \textup{(Normal form)}\; every formula is equivalent to a formula in
ADNF that is computable from it once its bound variables are renamed so that
no point variable is quantified inside a displacement whose parameter
contains it; the normal form has the same free variables and at most
quadratic size, and has the same quantifiers and bounded modalities, nested
in the same way and with the same parameters, as a negation normal form of
the renamed formula.
\end{enumerate}
\end{prop}
\begin{proof}
Parts (i)--(iv) follow from the laws of translation, displacement and the
bounded modalities in~\cite[Section~2]{zhang2024temporal}, since covariance holds. Part (v)
is the atomic-displacement normal form theorem proved there. Renaming
supplies the freshness required in (ii) when a displacement moves inwards
across a quantifier. For instance, at a point
$a$ the formula $(\boldBox_t\varphi)_u$
inspects the window $B(a+\eta(u),\eta(t))$, and $\boldBox_t(\varphi_u)$
inspects the set $B(a,\eta(t))+\eta(u)$; by covariance the two coincide. For
(iv), the point $a+\eta(u)$ lies at the distance $\norm{\eta(u)}$ from $a$,
hence in the window of length $\eta(t)+\norm{\eta(u)}$ at $a$.
\end{proof}

\noindent
The freshness condition in (ii), and hence the renaming in (v), is
necessary. If $p$ holds at the origin only and $\eta(x)\neq0$, then at the
origin $(\exists x\,p)_x$ fails: it tests $p$ at the point $\eta(x)$, whereas
$\exists x\,(p_x)$ holds. The normal form of (v) exposes the polarity of
atomic occurrences, which defines the enclosures of
Section~\ref{sec:enclosures}. It also brings displacements down to literals,
a property used in Section~\ref{sec:discussion}.

As in~\cite{zhang2024temporal}, the \emph{unbounded modalities} are defined by
quantification over a displacement: for a point variable $x$ that is fresh
for $\varphi$,
\[
\boldBox\varphi:=\forall x\,\varphi_x
\qquad\text{and}\qquad
\mdiamond\varphi:=\exists x\,\varphi_x .
\]
In a group, displacement can reach any point from any other. The two
modalities therefore read ``everywhere'' and ``somewhere''.

\begin{lem}[Unbounded modalities]\label{lem:unbounded}
For every interpretation $\alpha$, assignment $\eta$ and point $a$,
\[
(\alpha,\eta,a)\models\mdiamond\varphi
 \iff\sem{\varphi}_{\alpha,\eta}\neq\emptyset
\qquad\text{and}\qquad
(\alpha,\eta,a)\models\boldBox\varphi
 \iff\sem{\varphi}_{\alpha,\eta}=\Real^k .
\]
In particular, the truth of $\mdiamond\varphi$ and of $\boldBox\varphi$ does
not depend on the point of evaluation.
\end{lem}
\begin{proof}
As $x$ is fresh for $\varphi$, the formula $\varphi_x$ holds at $a$ under
$\eta[x\mapsto b]$ if and only if $\varphi$ holds at $a+b$ under $\eta$. As
$b$ ranges over $\Real^k$, so does $a+b$.
\end{proof}

A change of image coordinates should leave a specification intact, provided
that its metric and its displacements are transformed along with it. Let
$f(a):=Ma+c_0$ for an invertible real $k\times k$ matrix $M$ and a point
$c_0$, and let $G':=(M^{-1})^{\tp}GM^{-1}$, with norm $\norm{v}'$ and open
balls $B'(a,\ell)$. Then $\norm{Mv}'=\norm{v}$ for all $v$, and consequently
$f(B(a,\ell))=B'(f(a),\ell)$. For an interpretation $\alpha$ put
$\alpha':=\alpha\circ f^{-1}$; for a formula $\varphi$ let $\varphi'$ result
from $\varphi$ by replacing every point constant $c$ by $Mc$; and for an
assignment $\eta$ let $\eta'$ map every point variable $x$ to $M\eta(x)$ and
agree with $\eta$ on the length variables.

\begin{prop}[Change of coordinates]\label{prop:coordinates}
Let the entries of $M$ lie in $\Ralg$. Then $\varphi'$ is a formula, and
$(\alpha,\eta,a)\models\varphi$ in Euclidean space with the metric matrix $G$
if and only if $(\alpha',\eta',f(a))\models\varphi'$ in Euclidean space with
the metric matrix $G'$.
\end{prop}
\begin{proof}
The points $Mc$ have coordinates in $\Ralg$, so that $\varphi'$ is a formula.
We proceed by induction on $\varphi$. For an atomic proposition,
$p\in\alpha(a)$ if and only if $p\in\alpha'(f(a))$. The point denoted by an
affine term is transformed by $M$, since $M$ is linear, and
$f(a+b)=f(a)+Mb$; this gives the clause for displacement. The identity
$\norm{Mv}'=\norm{v}$ shows that a positive affine term denotes the same
length before and after the transformation, including terms with norms of
variables. The identity for balls and the bijectivity of $f$ give the clauses
for the bounded modalities. The map $M$ is a bijection of the domain of the
point variables, and the length domain is unchanged, which gives the clauses
for the quantifiers. The Boolean clauses are immediate.
\end{proof}

\noindent
Apart from its algebraic entries, the proof uses only that $M$ defines a
linear bijection $v\mapsto Mv$ with $\norm{Mv}'=\norm{v}$. Satisfaction is
invariant under a norm-preserving linear bijection followed by a translation;
the formula for $G'$ expresses this property in coordinates.
A positive-definite $G$ with entries in $\Ralg$ admits a factorization
$G=M^{\tp}M$ with $M$ invertible and with entries in $\Ralg$, for example by
Cholesky factorization. Then $G'=I$. For the effective metric
matrices considered below we may therefore reduce to the standard norm,
provided that the interpretation and the specification are transformed
together.

\section{Geometric Element Ensembles}\label{sec:gel}

An interpretation labels the points of a continuous space. A model checker
needs a finite description of these labels. For rational time, the labelled
interval ensembles of~\cite{qel} present atomic supports as finite unions
of intervals. Their spatial counterpart is the geometric element ensemble.

\subsection{Finite presentations of interpretations}\label{subsec:gels}
A \emph{presentation} is a quantifier-free formula $\Theta(y_1,\ldots,y_k)$
in the language of ordered rings with coefficients in $\Ralg$, that is, a
finite Boolean combination of polynomial equalities and inequalities. It
\emph{presents} the set $\{b\in\Real^k\mid\Real\models\Theta(b)\}$. A subset
of $\Real^k$ is \emph{semi-algebraic} if it is defined by such a Boolean
combination with real coefficients, and \emph{effective semi-algebraic} if it
has a presentation; a presentation is \emph{rational} if its coefficients lie
in $\Rat$. A presented set may be open, closed or neither, may be unbounded,
and may have dimension less than $k$.

\begin{defi}[Geometric element ensemble]\label{def:gel}
Let $\Prop$ be a finite nonempty set of atomic propositions. A
\emph{geometric element} of $\Real^k$ is an effective semi-algebraic subset
of $\Real^k$ given by a presentation. A \emph{geometric ensemble} is a finite
list $\mathcal{E}=(E_1,\ldots,E_n)$ of geometric elements, $n\geq0$, with
\emph{union} $|\mathcal{E}|:=E_1\cup\cdots\cup E_n$; we write $\GE_k$ for the
set of geometric ensembles of $\Real^k$. A \emph{geometric element ensemble}
(GEL) over $\Real^k$ is a map $\varepsilon\colon\Prop\to\GE_k$. It presents
the interpretation
\[
\alpha_\varepsilon\colon\Real^k\to2^{\Prop},\qquad
\alpha_\varepsilon(a):=\{p\in\Prop\mid a\in|\varepsilon(p)|\},
\]
and we write $(\varepsilon,\eta,a)\models\varphi$ for
$(\alpha_\varepsilon,\eta,a)\models\varphi$ and
$\sem{\varphi}_{\varepsilon,\eta}$ for the support
$\sem{\varphi}_{\alpha_\varepsilon,\eta}$, omitting $\eta$ for a closed
$\varphi$. A GEL is \emph{rational} if all its presentations are rational. A
\emph{ball ensemble} is a geometric ensemble whose elements are closed balls
$D(c,\rho)$ with $c\in\Ralg^k$ and $\rho\in\Ralg$, $\rho\geq0$, each
presented by the inequality $\inner{y-c}{y-c}\leq\rho^2$, and a
\emph{ball-GEL} is a GEL all of whose ensembles are ball ensembles.
\end{defi}

\noindent
By construction, $\sem{p}_\varepsilon=|\varepsilon(p)|$, and the empty list
presents an atomic proposition that holds nowhere. In the intended
applications a GEL describes one imaging examination. We call such a GEL a
\emph{record}. The semantics depends only on the union of the elements.
They may overlap, and replacing the list by another with the same union
changes no truth value. Unlike the canonical interval ensembles
of~\cite{zhang2024temporal}, a GEL has no distinguished decomposition into simple elements.
Equality of presented sets is nevertheless decidable: emptiness of their
symmetric difference is expressible in the theory of real closed
fields~\cite{tarski1951}.

\begin{exa}[A running example]\label{ex:running}
Let $G=I$ and $\Prop=\{L,V,O\}$, read as lesion, vessel and observed region.
The ball-GEL $\varepsilon_{\ast}$ with
\[
\varepsilon_{\ast}(L)=(D(0,0)),\qquad
\varepsilon_{\ast}(V)=(D(e_1,0)),\qquad
\varepsilon_{\ast}(O)=(D(0,2))
\]
is defined in every dimension $k$, and all its centres and radii are
rational. At the origin $L$ holds and $V$ fails, whereas the displaced
formula $V_{e_1}$ holds, since it tests $V$ at $0+e_1$. For a length
constant $r$, the formula $\mdiamond_rV$ holds there if and only if
$e_1\in B(0,r)$, that is, $r>1$, and $\boldBox_rO$ holds if and only if
$B(0,r)\subseteq D(0,2)$, that is, $r\leq2$. Hence
$\exists z\,(\mdiamond_zV\wedge\boldBox_zO)$ holds at the origin, witnessed
by every length in the interval $(1,2]$, and $\forall z\,\mdiamond_zV$
fails, refuted by every length $z\leq1$. As sets,
$\sem{\mdiamond_rV}_{\varepsilon_{\ast}}=B(e_1,r)$ and
$\sem{\boldBox_rO}_{\varepsilon_{\ast}}=D(0,2-r)$ for $r\leq2$.
\end{exa}

Rational boxes and polytopes are also geometric elements and present finite
pixel and voxel masks. Balls alone do not suffice: for $k\geq2$ no finite
union of balls equals a box. A ball contained in a closed half-space meets
its bounding hyperplane in at most one point, whereas a box facet is
infinite. Replacing a mask by a ball-GEL therefore calls for an enclosure (Section~\ref{sec:enclosures}). A change of coordinates, by
contrast, preserves the representation: for $c_0\in\Ralg^k$ in
Proposition~\ref{prop:coordinates}, substituting the affine expressions
$M^{-1}(y-c_0)$ for the variables of each presentation of $\varepsilon$
yields a GEL that presents $(\alpha_\varepsilon)'$.

The \emph{model-checking problem} for $\ELR$ asks whether
$(\varepsilon,a)\models\psi$, given a GEL $\varepsilon$ over $\Real^k$, a
point $a\in\Ralg^k$ and a closed formula $\psi$. Satisfaction itself remains
defined at real points under arbitrary interpretations.

\subsection{From the labels of an image to a record}\label{subsec:labels}
An acquisition labels a sampling grid. Its geometric interpretation specifies
what holds between samples: a union of grid cells is one choice, a
reconstructed surface or a union of balls another. Reconstructions that
agree at the stored samples may disagree on bounded universality. The
reconstruction convention is therefore part of the record. A query asserts
a geometric property: $\boldBox_rV$ tests for an inscribed ball in the
represented vessel. Reading it in terms of a centreline or a diameter
requires the definitions of Section~\ref{sec:vascular}.

\subsection{Observation policies and queries on a record}\label{subsec:observation}
Negation takes complements in $\Real^k$, so that the negation of an atomic
proposition with empty support holds everywhere. This closed-world reading
expresses anatomical absence only where the annotation is complete. An
atomic proposition $O$ for the observed region lets a query state how it
treats unobserved locations. For a formula $\varphi$ and a positive affine
term $t$, define
\[
\mdiamond^{O}_t\varphi:=\mdiamond_t(O\wedge\varphi),\qquad
\boldBox^{O}_t\varphi:=\boldBox_t(O\rightarrow\varphi),\qquad
\Full_t\varphi:=\boldBox_t(O\wedge\varphi).
\]
The first policy asks for an observed witness. The second, the
clipped-window policy, constrains the observed points of the window.
The third, the complete-window policy, requires complete observation of the
window and satisfaction of $\varphi$ throughout it. By
Proposition~\ref{prop:laws}(iii), it is equivalent to
$\boldBox_tO\wedge\boldBox_t\varphi$. The first two policies are dual,
$\boldBox^{O}_t\varphi\equiv\neg\mdiamond^{O}_t\neg\varphi$, and complete
observation entails both:
$\Full_t\varphi\models\boldBox^{O}_t\varphi$ and
$\Full_t\varphi\models\mdiamond^{O}_t\varphi$, the second because windows
are nonempty. Neither converse holds: if $O$ holds on a closed half-space
and $a$ is a point of its boundary, then at $a$ the formulas
$\boldBox^{O}_tO$ and $\mdiamond^{O}_tO$ hold, and $\Full_tO$, which is
equivalent to $\boldBox_tO$, fails, whatever length $t$ denotes.

An unbounded modality turns a local condition into a phenotype of the
record: $\mdiamond(O\wedge\theta)$ asserts that some observed location
satisfies $\theta$. By Lemma~\ref{lem:unbounded}, its truth is independent of
the point of evaluation. We evaluate such formulas at the origin. For a
\emph{database} $\mathcal{D}$, meaning a finite set of GELs over $\Real^k$,
and a closed formula $\chi$, the \emph{cohort} of $\chi$ is
$\Cohort_{\mathcal{D}}(\chi):=\{\varepsilon\in\mathcal{D}\mid(\varepsilon,0)\models\chi\}$.
Cohort discovery evaluates one closed formula separately on each record; the
geometries of different patients are never merged.

\section{Spatial Morphology of the Modal Operators}\label{sec:morphology}

The modal operators have a direct geometric reading. The closure question
has two levels: the class of single elements and the class of regions
presented by their ensembles. Balls admit a calculus
(Sections~\ref{subsec:morph} and~\ref{subsec:balls}). For unions, closure
depends on the operation and the dimension
(Sections~\ref{subsec:elementwise}--\ref{subsec:windows}). Evaluation
therefore requires a larger class of sets. The translation
theorem does not depend on Sections~\ref{subsec:topology}
and~\ref{subsec:cylinder}.

\subsection{Semantics as spatial morphology}\label{subsec:morph}
When truth at a point is replaced by the support of a formula, the
satisfaction clauses of Definition~\ref{def:sat} become operations on sets,
as they do for LINEs~\cite{zhang2024temporal}. For the windows considered here these are
the familiar operations of mathematical
morphology~\cite{matheron1975,serra1982}.

\begin{defi}[Morphological operations]\label{def:morph}
Let $S\subseteq\Real^k$, let $c$ be a point and $\ell$ a length. Define
\[
S_c:=\{a\mid a+c\in S\},\qquad
\mdiamond_\ell S:=\{a\mid B(a,\ell)\cap S\neq\emptyset\},\qquad
\boldBox_\ell S:=\{a\mid B(a,\ell)\subseteq S\};
\]
here the modal symbols denote operations on sets. The first operation is a
translation, $S_c=S-c$. In morphological terms,
$\mdiamond_\ell S=S+B(0,\ell)=\bigcup_{h\in B(0,\ell)}(S+h)$ is the
\emph{dilation} of $S$ by the open ball of radius $\ell$, and
$\boldBox_\ell S=\bigcap_{h\in B(0,\ell)}(S-h)$ is the \emph{erosion} of $S$
by this ball; the two are dual,
$\boldBox_\ell S=(\mdiamond_\ell(S^{c}))^{c}$.
\end{defi}

\begin{prop}[Compositional semantics; cf.~\cite{zhang2024temporal}]\label{prop:semantics}
Let $\alpha$ be an interpretation and $\eta$ an assignment, and write
$\sem{\varphi}$ for $\sem{\varphi}_{\alpha,\eta}$. For all formulas
$\varphi,\psi$, affine terms $u$ and positive affine terms $t$,
\[
\sem{\neg\varphi}=\sem{\varphi}^{c},\qquad
\sem{\varphi\wedge\psi}=\sem{\varphi}\cap\sem{\psi},\qquad
\sem{\varphi\vee\psi}=\sem{\varphi}\cup\sem{\psi},
\]
\[
\sem{\varphi_u}=\sem{\varphi}_{\eta(u)},\qquad
\sem{\mdiamond_t\varphi}=\mdiamond_{\eta(t)}\sem{\varphi},\qquad
\sem{\boldBox_t\varphi}=\boldBox_{\eta(t)}\sem{\varphi},
\]
\[
\sem{\exists x\,\varphi}_{\alpha,\eta}
  =\bigcup_{b\in\Real^k}\sem{\varphi}_{\alpha,\eta[x\mapsto b]},\qquad
\sem{\exists z\,\varphi}_{\alpha,\eta}
  =\bigcup_{\ell>0}\sem{\varphi}_{\alpha,\eta[z\mapsto\ell]},
\]
with intersections in place of unions for the universal quantifiers.
\end{prop}
\begin{proof}
The identities restate the clauses of Definition~\ref{def:sat}.
\end{proof}

\noindent
The parameter of a displacement translates the support by $-\eta(u)$,
because evaluation looks ahead by $\eta(u)$. The two bounded operations obey
the laws of an adjoint pair.

\begin{prop}[Morphological laws]\label{prop:morph}
Let $S,T\subseteq\Real^k$, and let $\ell,\ell'$ be lengths.
\begin{enumerate}[label=\textup{(\roman*)},leftmargin=2.2em]
\item Dilation preserves arbitrary unions, erosion preserves arbitrary
intersections, and both are monotone with respect to inclusion.
\item \textup{(Adjunction)}\; $\mdiamond_\ell S\subseteq T$ if and only if
$S\subseteq\boldBox_\ell T$.
\item \textup{(Composition)}\;
$\mdiamond_\ell\mdiamond_{\ell'}S=\mdiamond_{\ell+\ell'}S$ and
$\boldBox_\ell\boldBox_{\ell'}S=\boldBox_{\ell+\ell'}S$.
\item \textup{(Monotonicity in the length)}\; If $\ell\leq\ell'$, then
$\mdiamond_\ell S\subseteq\mdiamond_{\ell'}S$ and
$\boldBox_{\ell'}S\subseteq\boldBox_\ell S$.
\item \textup{(Opening and closing)}\;
$\mdiamond_\ell\boldBox_\ell S\subseteq S\subseteq\boldBox_\ell\mdiamond_\ell S$,
and the composites $\mdiamond_\ell\boldBox_\ell$ and
$\boldBox_\ell\mdiamond_\ell$ are monotone and idempotent.
\item \textup{(Distance)}\; $a\in\mdiamond_\ell S$ if and only if
$\dist(a,S)<\ell$, and $a\in\boldBox_\ell S$ if and only if
$\dist(a,S^{c})\geq\ell$.
\end{enumerate}
\end{prop}
\begin{proof}
(i) Dilation is a union of translates and erosion an intersection.
Translation commutes with unions and intersections. (ii) Both sides state
that $a+h\in T$ for all $a\in S$ and $h\in B(0,\ell)$. (iii) It suffices to
prove $B(0,\ell)+B(0,\ell')=B(0,\ell+\ell')$, since then
$\mdiamond_\ell\mdiamond_{\ell'}S=S+B(0,\ell')+B(0,\ell)=\mdiamond_{\ell+\ell'}S$.
The inclusion from left to right is the triangle inequality. Conversely, if
$\norm{v}<\ell+\ell'$, then $v$ is the sum of $\frac{\ell}{\ell+\ell'}v$ and
$\frac{\ell'}{\ell+\ell'}v$, whose norms are less than $\ell$ and $\ell'$.
The identity for erosion follows by two applications of duality,
$\boldBox_\ell\boldBox_{\ell'}S=(\mdiamond_\ell\mdiamond_{\ell'}(S^{c}))^{c}
=(\mdiamond_{\ell+\ell'}(S^{c}))^{c}=\boldBox_{\ell+\ell'}S$. (iv) If
$\ell\leq\ell'$, then $B(0,\ell)\subseteq B(0,\ell')$: a larger window offers
more witnesses and imposes more constraints. (v) The two inclusions are the
instances of (ii) for the pairs $(\boldBox_\ell S,S)$ and
$(S,\mdiamond_\ell S)$. Apply the first inclusion to
$\mdiamond_\ell\boldBox_\ell S$ and the second to
$\boldBox_\ell S$, following the latter by the monotone operation $\mdiamond_\ell$.
These give the two inclusions between $\mdiamond_\ell\boldBox_\ell S$ and
$\mdiamond_\ell\boldBox_\ell\mdiamond_\ell\boldBox_\ell S$. The argument
for the closing is dual. (vi) An infimum of distances is less
than $\ell$ exactly when some point of the set is at a distance less than
$\ell$; the second equivalence follows from the first, applied to $S^{c}$, by
duality.
\end{proof}

By Proposition~\ref{prop:semantics}, these laws yield equivalences and
entailments of $\ELR$: for positive affine terms $s,t$ and formulas
$\varphi$,
\[
\mdiamond_s\mdiamond_t\varphi\equiv\mdiamond_{s+t}\varphi,\qquad
\boldBox_s\boldBox_t\varphi\equiv\boldBox_{s+t}\varphi,\qquad
\mdiamond_t\boldBox_t\varphi\models\varphi\models\boldBox_t\mdiamond_t\varphi .
\]
The first two extend the composition law of dense time from~\cite{zhang2024temporal} to
space in arbitrary dimension. We call $\dist(a,S^{c})$ the \emph{clearance} of
$a$ in $S$. By part (vi), $\boldBox_t\varphi$ holds at $a$ exactly when the
clearance of $a$ in the support of $\varphi$ is at least the length denoted
by $t$; the inequality is weak because the window is open.

\subsection{Balls and collective coverage}\label{subsec:balls}
On a single ball the three operations have closed forms. The radius of a
closed ball may be $0$, whereas the syntax admits no window of length $0$.

\begin{prop}[Operations on a ball]\label{prop:ball}
Let $\rho\geq0$, let $\ell$ be a length, let $c,v$ be points, and let
$D:=D(c,\rho)$. Then
\[
D_v=D(c-v,\rho),\qquad
\mdiamond_\ell D=B(c,\rho+\ell),\qquad
\boldBox_\ell D=
 \begin{cases}
  D(c,\rho-\ell),&\rho\geq\ell,\\
  \emptyset,&\rho<\ell.
 \end{cases}
\]
For an open ball $B(c,\rho)$ with $\rho>0$, the dilation and the erosion are
given by the same right-hand sides.
\end{prop}
\begin{proof}
The first identity holds because translations are isometries. For the
dilation, if $b=b'+h$ with $\norm{b'-c}\leq\rho$ and $\norm{h}<\ell$, then
$\norm{b-c}<\rho+\ell$ by the triangle inequality. Conversely, let
$\mu:=\norm{b-c}<\rho+\ell$. If $\mu\leq\rho$, then $b\in D$. Otherwise put
$b':=c+\frac{\rho}{\mu}(b-c)$; then $\norm{b'-c}=\rho$ and
$\norm{b-b'}=\mu-\rho<\ell$. For $\rho=0$ this says that the dilation of
$\{c\}$ is the open ball of radius $\ell$.

For the erosion, fix a point $a$ and let $\mu:=\norm{a-c}$. We claim that
\[
\sup\{\norm{a+h-c}\mid\norm{h}<\ell\}=\mu+\ell
\]
and that the supremum is not attained. The bound
$\norm{a+h-c}\leq\mu+\norm{h}<\mu+\ell$ is the triangle inequality. For the
converse, choose a vector $w$ of norm $1$, equal to $(a-c)/\mu$ if $\mu>0$
and arbitrary otherwise; then $h=\lambda w$ with $0\leq\lambda<\ell$ gives
$\norm{a+h-c}=\mu+\lambda$, which tends to $\mu+\ell$ as $\lambda$ tends to
$\ell$. Now $B(a,\ell)\subseteq D$ if and only if every value
$\norm{a+h-c}$ is at most $\rho$, that is, if and only if
$\mu+\ell\leq\rho$.

For the open ball $B(c,\rho)$, containment of the window requires every
value to be less than $\rho$; since the supremum is not attained, this is
again the condition $\mu+\ell\leq\rho$. The argument for the dilation applies
with $\norm{b'-c}<\rho$: if $\rho\leq\mu<\rho+\ell$, choose $\rho'$ with
$\mu-\ell<\rho'<\rho$ and $\rho'\geq0$, and put
$b':=c+\frac{\rho'}{\mu}(b-c)$. The hypothesis $\rho>0$ is needed because an
open ball of radius $0$ is empty.
\end{proof}

The formulas record the dimension and boundary status of their outputs.
For a positive radius, $\boldBox_\rho D(c,\rho)=\{c\}$ is a single point, and dilation
turns a closed ball into an open one. For a region with several elements, the clearance of
Proposition~\ref{prop:morph}(vi) is measured in their union. It need not
equal the clearance in any one element.

\begin{exa}[Coverage can require several balls]\label{ex:coverage}
Let $G=I$ and
\[
D_{+}:=D\bigl(\tfrac34e_1,\tfrac54\bigr),\qquad
D_{-}:=D\bigl(-\tfrac34e_1,\tfrac54\bigr),\qquad
U:=D_{+}\cup D_{-}.
\]
For $b\in B(0,1)$, take the ball whose centre has the sign of $b_1$, and
either ball if $b_1=0$. Then
\[
\bigl|b\mp\tfrac34e_1\bigr|^2=|b|^2+\tfrac9{16}-\tfrac32|b_1|
<1+\tfrac9{16}=\tfrac{25}{16},
\]
so that $B(0,1)\subseteq U$. Neither ball contains this window: for
$1/2<\lambda<1$ the point $-\lambda e_1$ lies in the window and at the
distance $\lambda+3/4>5/4$ from the centre of $D_{+}$, and symmetrically for
$D_{-}$. Therefore
$0\in\boldBox_1(D_{+}\cup D_{-})\setminus
\bigl(\boldBox_1D_{+}\cup\boldBox_1D_{-}\bigr)$,
in agreement with $\boldBox_1D_{\pm}=D(\pm\tfrac34e_1,\tfrac14)$ from
Proposition~\ref{prop:ball}. The centres and radii are rational. Containment
of a window in a single element is thus a sound sufficient test for bounded
universality over a ball-GEL, but it is incomplete
(Figure~\ref{fig:operators}(c)). The clearance of the origin
in either ball alone is $5/4-3/4=1/2$. For $k\geq2$ its clearance in $U$ is
$1$: it is at least $1$ by the inclusion just proved, and for every
$\tau>0$ the point $(1+\tau)e_2$ lies outside both balls, because
$|(1+\tau)e_2\mp\tfrac34e_1|^2=(1+\tau)^2+\tfrac9{16}>\tfrac{25}{16}$.
\end{exa}

\subsection{Unions of balls: element-wise operations}\label{subsec:elementwise}
By Proposition~\ref{prop:ball}, the three operations preserve the class
consisting of the empty set and the open and closed balls, in arbitrary
dimension. On balls they reduce to arithmetic on a centre and a radius.
Their behaviour on finite unions differs.

\begin{prop}[Element-wise operations]\label{prop:elementwise}
Let $E_1,\ldots,E_n$ be balls of $\Real^k$, each open or closed, with centres
$c_i$ and radii $\rho_i$. Let $U:=E_1\cup\cdots\cup E_n$, let $\ell$ be a
length and $v$ a point.
\begin{enumerate}[label=\textup{(\roman*)},leftmargin=2.2em]
\item $U_v=\bigcup_i(E_i)_v$ and
$\mdiamond_\ell U=\bigcup_iB(c_i,\rho_i+\ell)$.
\item $\bigcup_i\boldBox_\ell E_i\subseteq\boldBox_\ell U$, where
$\boldBox_\ell E_i=D(c_i,\rho_i-\ell)$ if $\rho_i\geq\ell$ and
$\boldBox_\ell E_i=\emptyset$ otherwise.
\item Equality holds in \textup{(ii)} if the closures of $E_1,\ldots,E_n$
are pairwise disjoint.
\item If $\rho_i=\dist(c_i,U^{c})$, then $c_i\in\boldBox_\ell U$ if and only
if $\ell\leq\rho_i$.
\end{enumerate}
\end{prop}
\begin{proof}
Translation and dilation preserve unions
(Proposition~\ref{prop:morph}(i)). Proposition~\ref{prop:ball} evaluates
them on the elements, proving (i). Erosion is monotone, so that
$\boldBox_\ell E_i\subseteq\boldBox_\ell U$ for each $i$, and
Proposition~\ref{prop:ball} gives the stated form; this is (ii). For (iii),
let $B(a,\ell)\subseteq U$. The intersections of the window with the closures
of $E_1,\ldots,E_n$ are relatively closed in the window, pairwise disjoint,
and cover it. The window is connected, so that exactly one of them is
nonempty, say the one with index $i$. A point of the window then lies in $U$
and in no $E_j$ with $j\neq i$, hence in $E_i$; thus $a\in\boldBox_\ell E_i$.
Statement (iv) is Proposition~\ref{prop:morph}(vi) at the point $c_i$.
\end{proof}

\noindent
An element with $\rho_i=\dist(c_i,U^{c})$ is called \emph{tight}. Part (i)
evaluates formulas built from atomic propositions by disjunction,
displacement and bounded existence with closed parameters. Over a ball-GEL,
their supports remain finite unions of balls. For bounded
universality, part (ii) gives a certified inner bound, which can be strict
(Example~\ref{ex:coverage}), and parts (iii) and (iv) identify two
cases in which it gives the answer.

\subsection{Closure of finite unions of balls: the line and the higher
dimensions}\label{subsec:closure}
Write $\mathfrak{B}_k$ for the class of finite unions of balls of $\Real^k$,
each open or closed; the empty union is the empty set. By
Proposition~\ref{prop:elementwise}(i), $\mathfrak{B}_k$ is closed under
union, translation and dilation. Closure under intersection and erosion
depends on the dimension.

\begin{prop}[Closure of finite unions of balls]\label{prop:closure}
\begin{enumerate}[label=\textup{(\roman*)},leftmargin=2.2em]
\item Let $k=1$. Then $\mathfrak{B}_1$ is the class of finite unions of
bounded intervals. It is closed under intersection and under erosion, and the
erosion of a member of $\mathfrak{B}_1$ is the union of the erosions of its
connected components.
\item Let $k\geq2$. Then $\mathfrak{B}_k$ is closed neither under
intersection nor under erosion. For every orthonormal pair $(w_1,w_2)$,
neither the lens $D(0,1)\cap D(w_1,1)$ nor the erosion $\boldBox_1U$ of the
set $U:=D(\frac34w_1,\frac54)\cup D(-\frac34w_1,\frac54)$ belongs to
$\mathfrak{B}_k$; for $G=I$ and $w_1=e_1$ this is the set $U$ of
Example~\ref{ex:coverage}.
\end{enumerate}
\end{prop}
\begin{proof}
(i) For $k=1$ and $\norm{v}=\gamma|v|$, the balls are the intervals
$B(c,\rho)=(c-\rho/\gamma,c+\rho/\gamma)$ and
$D(c,\rho)=[c-\rho/\gamma,c+\rho/\gamma]$. A bounded interval is the union of
its interior, an open ball if it is nonempty, and of those of its endpoints
that it contains, which are closed balls of radius $0$. Hence
$\mathfrak{B}_1$ is the class of finite unions of bounded intervals. Two
bounded intervals intersect in a bounded interval or in the empty set, and
intersection distributes over union; this gives closure under intersection.
For erosion, note that the finite unions of intervals, bounded or unbounded,
form a Boolean algebra of subsets of the line, and that this algebra is
closed under dilation, since the dilation of an interval is an interval and
dilation preserves unions (Proposition~\ref{prop:morph}(i)). For
$S\in\mathfrak{B}_1$ the set
$\boldBox_\ell S=(\mdiamond_\ell(S^{c}))^{c}$ therefore belongs to the
algebra; it is contained in $S$, hence bounded, and so belongs to
$\mathfrak{B}_1$. Finally, a window is connected, so that
$B(a,\ell)\subseteq S$ if and only if $B(a,\ell)$ lies in one connected
component of $S$; this last argument applies to every subset of $\Real^k$,
in every dimension.

(ii) Let $(w_1,w_2)$ be an orthonormal pair. Both sets are closed: the lens
is an intersection of closed sets, and
$\boldBox_1U=\{a\mid\dist(a,U^{c})\geq1\}$ by
Proposition~\ref{prop:morph}(vi), where the distance from a fixed set is a
continuous function. A closed set that contains a ball with the centre $c$
and the radius $\rho$, open or closed, contains the closure $D(c,\rho)$ of
that ball.

\emph{The lens.} Let $K:=D(0,1)\cap D(w_1,1)$, and suppose that
$K=E_1\cup\cdots\cup E_m$ with balls $E_j$, each open or closed, with the
centres $c_j$ and the radii $\rho_j$. Then
$D(c_j,\rho_j)\subseteq K\subseteq D(0,1)$, so that
$\norm{c_j}+\rho_j\leq1$ by Proposition~\ref{prop:ball}. Let
$b\in D(c_j,\rho_j)$ with $\norm{b}=1$. Then
\[
1=\norm{b}\leq\norm{c_j}+\norm{b-c_j}\leq\norm{c_j}+\rho_j\leq1 ,
\]
so that equality holds throughout. Here $c_j\neq0$, since otherwise
$\rho_j=1$ and $D(c_j,\rho_j)=D(0,1)$ would contain the point $-w_1$, which
lies outside $D(w_1,1)$. The triangle inequality is an equality for $c_j$
and $b-c_j$; in the norm of an inner product this forces
$b-c_j=\lambda c_j$ with $\lambda\geq0$, whence $b=c_j/\norm{c_j}$. Each of
the balls $D(c_j,\rho_j)$ thus contains at most one point of norm $1$,
whereas $K$ contains infinitely many: the points $\lambda w_1+\mu w_2$ with
$\lambda^2+\mu^2=1$ and $\lambda\geq\frac12$ have the distance
$(2-2\lambda)^{1/2}\leq1$ from $w_1$.

\emph{The erosion.} Let $K:=\boldBox_1U$. We use three facts.

(a) For $0\leq\lambda\leq3/4$, the point $\lambda w_1$ belongs to $K$.
Indeed, let $b\in B(\lambda w_1,1)$, and write $b_1:=\inner{b}{w_1}$,
$\mu:=3/4-\lambda\geq0$ and
$\nu:=3/4+\lambda>0$, so that $\mu\nu=9/16-\lambda^2$. If $b_1\geq0$, then
$-(b_1-\lambda)\leq\lambda$ and
\[
\norm{b-\tfrac34w_1}^2
 =\norm{b-\lambda w_1}^2-2\mu(b_1-\lambda)+\mu^2
 <1+2\lambda\mu+\mu^2=1+\mu\nu=\tfrac{25}{16}-\lambda^2 .
\]
If $b_1<0$, then $b_1-\lambda<-\lambda$ and
\[
\norm{b+\tfrac34w_1}^2
 =\norm{b-\lambda w_1}^2+2\nu(b_1-\lambda)+\nu^2
 <1-2\lambda\nu+\nu^2=1+\mu\nu=\tfrac{25}{16}-\lambda^2 .
\]
In both cases $b\in U$.

(b) $K$ is disjoint from $B(w_2,1)$ and from $B(-w_2,1)$. Indeed, if
$\norm{a-w_2}<1$, then $\norm{a-(1+\tau)w_2}<1$ for some $\tau>0$, and
$(1+\tau)w_2\notin U$, because
$\norm{(1+\tau)w_2\mp\tfrac34w_1}^2=(1+\tau)^2+\tfrac9{16}>\tfrac{25}{16}$,
as in Example~\ref{ex:coverage}; hence $B(a,1)\not\subseteq U$. The case of
$-w_2$ is symmetric.

(c) A closed ball $E=D(c,\rho)$ with $\rho>0$ and $0\in E$ meets
$B(w_2,1)\cup B(-w_2,1)$. Indeed, let $c_2:=\inner{c}{w_2}$, and put $v:=c$
if $c_2\neq0$ and $v:=c+w_2$ if $c_2=0$. Then $v_2:=\inner{v}{w_2}\neq0$ and
$\inner{v}{c}=\norm{c}^2$, so that for $\lambda>0$
\[
\norm{\lambda v-c}^2=\norm{c}^2+\lambda\bigl(\lambda\norm{v}^2-2\norm{c}^2\bigr)
\qquad\text{and}\qquad
\norm{\lambda v\mp w_2}^2=1+\lambda\bigl(\lambda\norm{v}^2\mp2v_2\bigr).
\]
Since $0\in E$, we have $\norm{c}\leq\rho$. If $\norm{c}=\rho$, then
$c\neq0$ and the first expression is at most $\rho^2$ for
$\lambda\leq2\norm{c}^2/\norm{v}^2$; if $\norm{c}<\rho$, it is at most
$\rho^2$ for all sufficiently small $\lambda$. With the sign chosen so that
$\mp v_2=-|v_2|$, the second expression is less than $1$ as soon as
$\lambda<2|v_2|/\norm{v}^2$. Hence, for all sufficiently small $\lambda>0$,
the point $\lambda v$ lies in $E$ and in one of the two open balls.

Now suppose that $K=E_1\cup\cdots\cup E_m$ with balls $E_j$, each open or
closed. By (a) the points $\frac1nw_1$, $n\geq2$, belong to $K$, so that one
ball $E_j$ contains infinitely many of them. This ball has a positive radius
$\rho$, and the closed ball $D(c,\rho)$ about its centre $c$ lies in $K$ and
contains their limit $0$. By (c) it meets $B(w_2,1)\cup B(-w_2,1)$, which
contradicts (b).
\end{proof}

\begin{figure}[htbp]
\centering
\begin{tikzpicture}[font=\small,x=1.12cm,y=1.12cm]
\begin{scope}
\fill[inkblue!15] (0.75,0) circle (1.25) (-0.75,0) circle (1.25);
\draw[inkblue] (0.75,0) circle (1.25) (-0.75,0) circle (1.25);
\draw[dashed,inkred] (0,0) circle (1);
\fill (0,0) circle (1.6pt) node[below=1pt]{\scriptsize $0$};
\fill[inkblue] (0.75,0) circle (1.2pt) (-0.75,0) circle (1.2pt);
\node[inkblue] at (1.45,0.55){$D_{+}$};
\node[inkblue] at (-1.45,0.55){$D_{-}$};
\draw[inkred,thin] (0.34,0.94)--(0.62,1.52);
\node[inkred,right=-1pt] at (0.62,1.55){\scriptsize $B(0,1)$};
\node at (0,-2.45){(a) $U=D_{+}\cup D_{-}$ and the window $B(0,1)$};
\end{scope}
\begin{scope}[shift={(5.9,0)}]
\draw[inkblue!45] (0.75,0) circle (1.25) (-0.75,0) circle (1.25);
\draw[dashed,inkred!70] (0,1) circle (1) (0,-1) circle (1);
\fill[inkblue!45] (0.6,-0.2) arc[start angle=233.13,end angle=486.87,radius=0.25]
   arc[start angle=-53.13,end angle=-126.87,radius=1]
   arc[start angle=53.13,end angle=306.87,radius=0.25]
   arc[start angle=126.87,end angle=53.13,radius=1] -- cycle;
\draw[inkblue] (0.6,-0.2) arc[start angle=233.13,end angle=486.87,radius=0.25]
   arc[start angle=-53.13,end angle=-126.87,radius=1]
   arc[start angle=53.13,end angle=306.87,radius=0.25]
   arc[start angle=126.87,end angle=53.13,radius=1] -- cycle;
\fill (0,0) circle (1.6pt);
\fill[inkred!70] (0,1) circle (1.1pt) (0,-1) circle (1.1pt);
\node[inkred!70] at (0,1.45){\scriptsize $B(e_2,1)$};
\node[inkred!70] at (0,-1.45){\scriptsize $B(-e_2,1)$};
\node at (0,-2.45){(b) the erosion $\boldBox_1U$};
\end{scope}
\end{tikzpicture}
\caption{Erosion of a union of two balls in the plane
(Proposition~\ref{prop:closure}(ii) for $G=I$ and the pair $(e_1,e_2)$).
(a)~The window $B(0,1)$ lies in the union $U$ of the closed balls $D_{+}$ and
$D_{-}$ and in neither of them. (b)~The erosion $\boldBox_1U$ (shaded)
consists of the balls $\boldBox_1D_{\pm}$ of radius $1/4$ and of two cusps
that meet at the origin, where the set is pinched between the open unit
balls about $e_2$ and $-e_2$ (dashed).}
\label{fig:closure}
\end{figure}
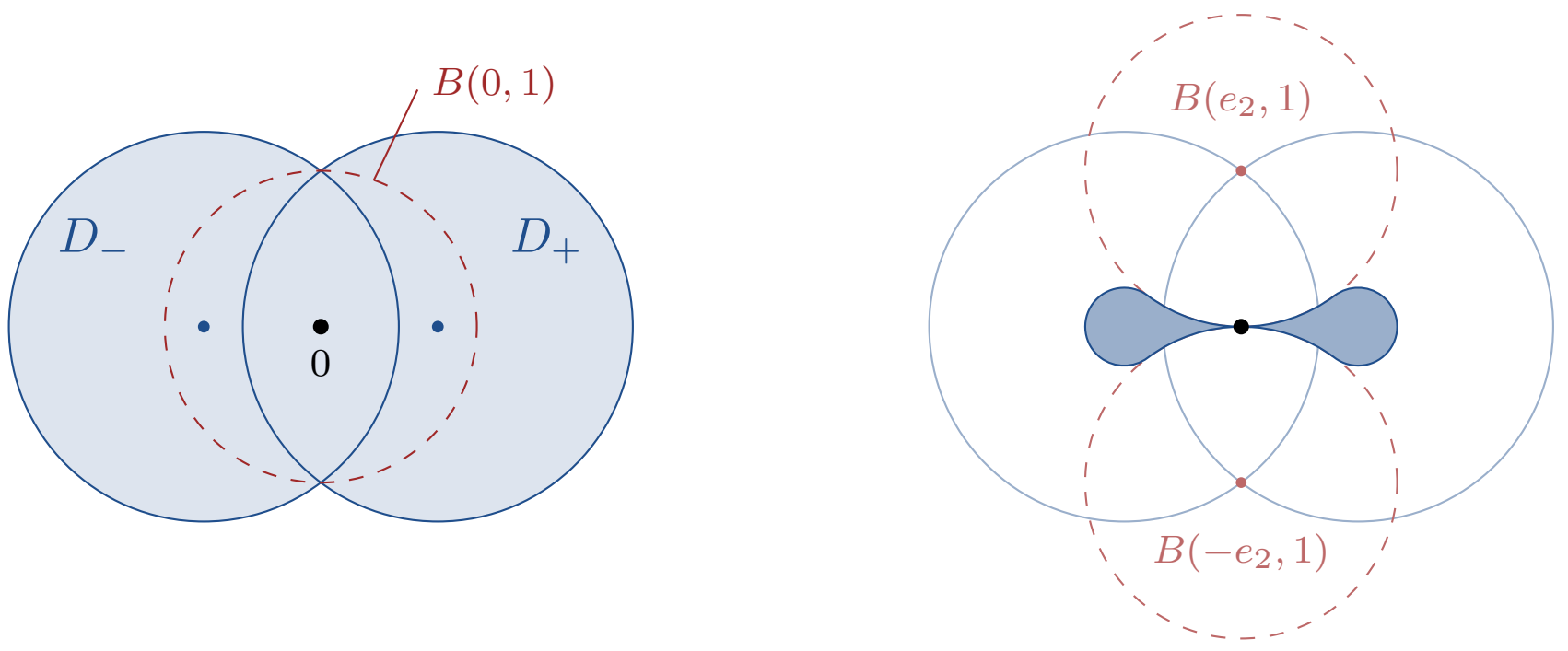

The shape of $\boldBox_1U$ explains the obstruction
(Figure~\ref{fig:closure}). Facts (a) and (b) place a segment through the
origin inside the eroded set and exclude two open unit balls tangent there.
This leaves no room for a ball of positive radius through the origin. For
$k\geq2$, an exact procedure cannot in general return a finite union of
balls. On the line, part (i) extends to the
interval ensembles of~\cite{zhang2024temporal}: finite unions of intervals, bounded or
unbounded, are closed under all logical operations, complement included.

\subsection{Nested, mixed and subdivided windows}\label{subsec:windows}
By Proposition~\ref{prop:morph}(iii), nesting two windows of the same kind
adds their lengths in arbitrary dimension. Mixed modalities and the
assembly of a window from smaller ones behave less uniformly.

\begin{prop}[Mixed and subdivided windows]\label{prop:windows}
Let $s,t$ be positive affine terms, $\varphi$ a formula, $p$ an atomic
proposition, and $r$ a length constant.
\begin{enumerate}[label=\textup{(\roman*)},leftmargin=2.2em]
\item $\mdiamond_s\boldBox_t\varphi\models\boldBox_t\mdiamond_s\varphi$, and
the converse entailment fails.
\item Let $k\geq2$. No finite family of open balls of radii less than $r$ has
the union $B(0,r)$. Consequently, for no point constants $c_1,\ldots,c_m$ and
length constants $r_1,\ldots,r_m<r$ is the equivalence
$\mdiamond_rp\equiv\bigvee_{j}(\mdiamond_{r_j}p)_{c_j}$ valid.
\item Let $k=1$, let $r'$ be a length constant with $r/2<r'<r$, and let $c$
be a point constant with $\norm{c}=r-r'$. Then
$\mdiamond_r\varphi\equiv(\mdiamond_{r'}\varphi)_{c}\vee(\mdiamond_{r'}\varphi)_{-c}$
and
$\boldBox_r\varphi\equiv(\boldBox_{r'}\varphi)_{c}\wedge(\boldBox_{r'}\varphi)_{-c}$.
\end{enumerate}
\end{prop}
\begin{proof}
(i) Write $s$ and $t$ also for the lengths that the two terms denote. Let
$b\in B(a,s)$ with $B(b,t)\subseteq\sem{\varphi}$, and let $b'\in B(a,t)$.
The point $w:=b+b'-a$ satisfies $\norm{w-b}=\norm{b'-a}<t$ and
$\norm{w-b'}=\norm{b-a}<s$; hence $\varphi$ holds at the point $w$ of
$B(b',s)$. For the converse, let $p$ hold at the origin only, and let $s,t$
be length constants with $t\leq s$. At the origin
$\boldBox_t\mdiamond_sp$ holds, since
$B(0,t)\subseteq B(0,s)=\sem{\mdiamond_sp}$, whereas $\boldBox_tp$ holds
nowhere.

(ii) Suppose that $B(0,r)=\bigcup_{j\leq m}B(c_j,r_j)$ with $r_j<r$ for all
$j$. Each ball of the family lies in $B(0,r)$, so that
$\norm{c_j}+r_j\leq r$ by Proposition~\ref{prop:ball}. Taking closures gives
$D(0,r)=\bigcup_jD(c_j,r_j)$. Let $\norm{b}=r$ and $b\in D(c_j,r_j)$. Then
\[
r=\norm{b}\leq\norm{c_j}+\norm{b-c_j}\leq\norm{c_j}+r_j\leq r ,
\]
so that equality holds throughout: $\norm{c_j}=r-r_j>0$, and the triangle
inequality is an equality for $c_j$ and $b-c_j$. In the norm of an inner
product this forces $b-c_j=\lambda c_j$ with $\lambda\geq0$, whence
$b=rc_j/\norm{c_j}$. A ball of the family thus accounts for at most one
point of the sphere $\{b\mid\norm{b}=r\}$. The sphere is infinite for $k\geq2$,
giving a contradiction. For the logical statement, evaluate both sides at the origin
under the interpretations in which $p$ holds at exactly one point $b$: the
left-hand side holds if and only if $b\in B(0,r)$, and the right-hand side
if and only if $b\in\bigcup_jB(c_j,r_j)$. Validity would make the two sets
equal.

(iii) Let $w:=c/\norm{c}$, so that the points of the line are the points
$a+\tau w$ with $\tau\in\Real$, and $\norm{\tau w}=|\tau|$. At a point $a$,
the two displaced modalities inspect the windows $B(a+c,r')$ and
$B(a-c,r')$, which consist of the points $a+\tau w$ with
$r-2r'<\tau<r$ and with $-r<\tau<2r'-r$. The two intervals overlap because
$2r'>r$. Their union is the interval $-r<\tau<r$, which describes the
window $B(a,r)$.
\end{proof}

Part (ii) distinguishes spatial windows from those of linear time. The
temporal window splits as $[a,a+s+t)=[a,a+s)\cup[a+s,a+s+t)$ by the additive
decomposition law of~\cite{zhang2024temporal}. That work also observes that two balls of
radii $s$ and $t$ do not cover a ball of radius $s+t$ when $k\geq2$.
Part (ii) rules out a finite union of smaller balls. A window can thus be
composed through nesting but not replaced by a finite subdivision.

\subsection{Derived topological operators and locality}\label{subsec:topology}
Quantification over lengths yields topological operators without new
primitives. For a length variable $z$ that is fresh for $\varphi$, define
\[
\Intr\,\varphi:=\exists z\,\boldBox_z\varphi,\qquad
\Clo\,\varphi:=\forall z\,\mdiamond_z\varphi,\qquad
\partial\varphi:=\Clo\,\varphi\wedge\neg\,\Intr\,\varphi .
\]

\begin{prop}[Interior, closure and boundary]\label{prop:topology}
For every interpretation and assignment, the support of $\Intr\,\varphi$ is
the interior of the support of $\varphi$, the support of $\Clo\,\varphi$ is
its closure, and the support of $\partial\varphi$ is its boundary.
\end{prop}
\begin{proof}
A point is interior to a set exactly when some open ball about it lies in
the set, and it belongs to the closure exactly when every open ball about it
meets the set. The balls $B(a,\ell)$ form a base of the topology of $\Real^k$
(Section~\ref{subsec:mm}), and the clauses for $\exists z\,\boldBox_z$ and
$\forall z\,\mdiamond_z$ state these two conditions. The boundary is the
closure minus the interior.
\end{proof}

The derived operators express queries about boundaries
(Section~\ref{subsec:budget}). A locality argument shows that the fragment
without variables cannot express connectedness. A formula is
\emph{constant} if
it is quantifier-free and all its parameters are closed terms. For a
constant formula we define the \emph{observation radius}
$\rad(\varphi)\geq0$ by $\rad(p):=0$, $\rad(\neg\varphi):=\rad(\varphi)$,
$\rad(\varphi\wedge\psi):=\rad(\varphi\vee\psi):=\max\{\rad(\varphi),\rad(\psi)\}$,
\[
\rad(\varphi_u):=\norm{u}+\rad(\varphi)
\qquad\text{and}\qquad
\rad(\mdiamond_t\varphi):=\rad(\boldBox_t\varphi):=t+\rad(\varphi),
\]
where the closed terms $u$ and $t$ stand for the point and the length that
they denote. Two interpretations $\alpha,\alpha'$ \emph{agree on} a set
$S\subseteq\Real^k$ if $\alpha(b)=\alpha'(b)$ for every $b\in S$.

\begin{prop}[Finite observation radius]\label{prop:radius}
Let $\varphi$ be a constant formula and $a$ a point. If two interpretations
agree on $D(a,\rad(\varphi))$, then $\varphi$ holds at $a$ under both or
under neither.
\end{prop}
\begin{proof}
By induction on $\varphi$, for all points $a$ simultaneously. The atomic and
Boolean cases are immediate. A displaced formula $\varphi_u$ holds at $a$
exactly when $\varphi$ holds at $a+u$. The triangle inequality gives
$D(a+u,\rad(\varphi))\subseteq D(a,\norm{u}+\rad(\varphi))$,
so the induction hypothesis applies at $a+u$. For a bounded
modality with the parameter $t$, every point $b$ of the window satisfies
$d(a,b)<t$, hence $D(b,\rad(\varphi))\subseteq D(a,t+\rad(\varphi))$. The
induction hypothesis applies at every point of the window, and the two
interpretations give the same truth value to $\mdiamond_t\varphi$ and to
$\boldBox_t\varphi$ at $a$.
\end{proof}

\begin{cor}[Connectedness]\label{cor:connected}
Let $p\in\Prop$ and $a\in\Ralg^k$. There is no constant formula $\varphi$
such that, for every ball-GEL $\varepsilon$ over $\Real^k$,
$(\varepsilon,a)\models\varphi$ if and only if $\sem{p}_\varepsilon$ is
connected.
\end{cor}
\begin{proof}
Let $\varphi$ be a constant formula. Let $\varepsilon$ have
$\varepsilon(p)=(D(a,1))$ and the empty list for every other atomic
proposition, and let $\varepsilon'$ differ from $\varepsilon$ by one further
element $D(a',1)$ of $p$, where $a'\in\Ralg^k$ and
$d(a,a')>\rad(\varphi)+2$. The two GELs agree on $D(a,\rad(\varphi))$, so
that $\varphi$ has the same truth value at $a$ in both, by
Proposition~\ref{prop:radius}. Yet $\sem{p}_\varepsilon$ is connected and
$\sem{p}_{\varepsilon'}$ is not.
\end{proof}

The restriction to constant formulas is essential. Point quantification
removes the fixed observation radius on which the argument rests. The
corollary therefore establishes no undefinability result for the full logic.

\subsection{Cylindrical transfer between dimensions}\label{subsec:cylinder}
Comparing a planar record with a volume requires a choice of what holds
away from the plane. Let $1\leq k'<k$. Embedding the supports of $\Real^{k'}$
in a subspace of $\Real^k$ destroys bounded universality: windows of
$\Real^k$ also inspect points outside the subspace. Cylindrical extension
continues the labels unchanged along the added coordinates.

Let $\pi\colon\Real^k\to\Real^{k'}$ be the projection onto the first $k'$
coordinates, and let the metric matrices $G$ of $\Real^k$ and $G'$ of
$\Real^{k'}$ be compatible: $G$ is block diagonal with the blocks $G'$ and
$H$, where $H$ is positive definite. We write $\norm{v}'$ and $B'(a,\ell)$
for the norm and the open balls of $\Real^{k'}$, and $\norm{v}_H$ for the
norm of $H$. For $v=(v',v'')\in\Real^{k'}\times\Real^{k-k'}$ this gives
\begin{equation}\label{eq:pythagoras}
\norm{v}^2=(\norm{v'}')^2+\norm{v''}_H^2 .
\end{equation}
The \emph{cylindrical lift} of an interpretation $\alpha$ over $\Real^{k'}$
is the interpretation $\alpha^{\uparrow}:=\alpha\circ\pi$ over $\Real^k$,
whose supports are the cylinders
$\sem{p}_{\alpha^{\uparrow}}=\pi^{-1}(\sem{p}_\alpha)$. If $\alpha$ is
presented by a GEL $\varepsilon$, then $\alpha^{\uparrow}$ is presented by
the GEL $\varepsilon^{\uparrow}$ that keeps every presentation
$\Theta(y_1,\ldots,y_{k'})$ and reads it over the variables
$y_1,\ldots,y_k$. The lift $\varphi^{\uparrow}$ of a formula $\varphi$ of
$\EL(\Real^{k'})$ replaces each point constant $c$ by $(c,0)$ and leaves
variables, length constants and all operators unchanged; $u^{\uparrow}$ and
$t^{\uparrow}$ denote the lifts of terms. For an assignment $\eta$ over
$\Real^k$, the assignment $\pi\eta$ over $\Real^{k'}$ maps $x$ to
$\pi(\eta(x))$ and $z$ to $\eta(z)$.

\begin{thm}[Cylindrical preservation]\label{thm:cylinder}
Let $1\leq k'<k$, and let $\varphi$ be a formula of $\EL(\Real^{k'})$ in
which every norm $\norm{u}$ that occurs in a positive affine term has a
closed term $u$. Then, for every interpretation $\alpha$ over $\Real^{k'}$,
every assignment $\eta$ over $\Real^k$ and every point $a$ of $\Real^k$,
\[
(\alpha^{\uparrow},\eta,a)\models\varphi^{\uparrow}
\quad\Longleftrightarrow\quad
(\alpha,\pi\eta,\pi a)\models\varphi .
\]
Equivalently,
$\sem{\varphi^{\uparrow}}_{\alpha^{\uparrow},\eta}
=\pi^{-1}\bigl(\sem{\varphi}_{\alpha,\pi\eta}\bigr)$.
\end{thm}
\begin{proof}
We use three facts. (a) For an affine term $u$,
$\pi(\eta(u^{\uparrow}))=(\pi\eta)(u)$: this holds for constants and
variables by definition, and for sums because $\pi$ is
linear. (b) For a positive affine term $t$ of $\varphi$,
$\eta(t^{\uparrow})=(\pi\eta)(t)$: length constants and variables are
unchanged, and a norm in $t$ has a closed argument $u$, whose lift denotes
the point $(u,0)$ with $\norm{(u,0)}=\norm{u}'$ by~\eqref{eq:pythagoras}.
(c) For a point $a$ of $\Real^k$ and a length $\ell$,
$\pi(B(a,\ell))=B'(\pi a,\ell)$. Indeed, by~\eqref{eq:pythagoras} the
projection does not increase the norm of $b-a$, and a point $b'$ of the
right-hand side is the image of $(b',a'')$, which lies at the same distance
from $a=(a',a'')$.

The equivalence is proved by induction on $\varphi$, for all $a$ and $\eta$
simultaneously. For an atomic proposition, $p\in\alpha^{\uparrow}(a)$ if and
only if $p\in\alpha(\pi a)$. The Boolean cases are immediate. For a
displacement, the induction hypothesis applies at the point
$a+\eta(u^{\uparrow})$, and
$\pi(a+\eta(u^{\uparrow}))=\pi a+(\pi\eta)(u)$ by (a). For
$\mdiamond_t\psi$, let $\ell:=\eta(t^{\uparrow})=(\pi\eta)(t)$, using (b). By
the induction hypothesis, the truth of $\psi^{\uparrow}$ at a point $b$
depends only on $\pi b$. Hence $\psi^{\uparrow}$ holds somewhere in
$B(a,\ell)$ if and only if $\psi$ holds somewhere in the projection of this
ball, which is $B'(\pi a,\ell)$ by (c). The case of $\boldBox_t\psi$ is the
same with ``everywhere'' in place of ``somewhere''. For
$\exists x\,\psi$, note that
$\pi(\eta[x\mapsto b])=(\pi\eta)[x\mapsto\pi b]$ and that $\pi b$ ranges over
$\Real^{k'}$ as $b$ ranges over $\Real^k$; the induction hypothesis for the
variant assignments gives the claim. The length quantifiers are treated in
the same way because the length domain is unchanged. The universal
quantifiers are dual.
\end{proof}

The hypotheses enter at specific steps of the argument. Without block
diagonality, $B(0,\ell)$ projects to the ball of radius $\ell$ for the Schur
complement $G'-NH^{-1}N^{\tp}$, where $N$ is the off-diagonal block of $G$.
This differs from $G'$ whenever $N\neq0$, so fact (c) fails. Norms of
variables obstruct fact (b), since in general $\norm{v}\neq\norm{\pi v}'$.
The theorem transfers a specification under a stated geometric extension.
For $k'=2$ and $k=3$, this is a stack of identical slices. It supports no
inference about an arbitrary volumetric structure from one planar section.

\section{Exact Model Checking}\label{sec:modelchecking}

The identities of Proposition~\ref{prop:semantics} describe how supports
compose. They do not yet give an algorithm: a length may depend on a
variable bound outside several operators. For rational time, a translation
into linear arithmetic resolves this difficulty~\cite{qel}. In dimension
$k\geq2$ balls are defined by quadratic inequalities. The appropriate target
is therefore the first-order theory of real closed fields, $\RCF$, in the
language of ordered rings. The translation of
Section~\ref{subsec:translation} is the main result of the paper. The
remainder of the section develops its consequences.

\subsection{The standard translation}\label{subsec:translation}
Fix a GEL $\varepsilon$ over $\Real^k$. For each atomic proposition $p$, let
$\Theta_p(y)$ be the disjunction of the presentations of the elements of
$\varepsilon(p)$; this presents $\sem{p}_\varepsilon$, with the empty
disjunction interpreted as false. The coefficients of the GEL, the constants
of a formula and the coordinates of a point of evaluation are real algebraic
numbers. They admit exact computation~\cite{bpr2006}, with rational data
as the principal case.

The arithmetic language is the language of ordered rings with constants for
the elements of $\Ralg$. Each point variable $x$ of the logic is read as a
$k$-tuple $(x_1,\ldots,x_k)$ of arithmetic variables, its
\emph{coordinates}. A length variable becomes a single arithmetic variable.
An assignment gives these arithmetic variables real values. An
\emph{evaluation tuple} is a $k$-tuple $e$ of polynomial terms of the
arithmetic language. An affine term $u$ translates into the $k$-tuple
$\widehat{u}$ of polynomials of degree at most $1$ that is obtained by
reading its additions coordinatewise, so that $\widehat{u}$ has the value
$\eta(u)$ under $\eta$.

Norms prevent the direct translation of a positive affine term into a
polynomial in the coordinates. For a positive affine term $t$, let
$\norm{u_1},\ldots,\norm{u_m}$ be the norms occurring in it, $m\geq0$, and
introduce fresh arithmetic variables $h=(h_1,\ldots,h_m)$. Write
$\widehat{t}$ for the polynomial of degree at most $1$ obtained from $t$ by
replacing each norm $\norm{u_i}$ by $h_i$. The \emph{norm constraint} of $t$
relates these auxiliary variables to the coordinates of the points,
\[
N_t(h):=\bigwedge_{i=1}^{m}\bigl(h_i\geq0\wedge
h_i^2=\inner{\widehat{u}_i}{\widehat{u}_i}\bigr).
\]

\begin{lem}[Norm constraints]\label{lem:norms}
For every positive affine term $t$ and every assignment $\eta$, exactly one
tuple of reals satisfies $N_t$ under $\eta$, namely the tuple of the norms
$\norm{\eta(u_i)}$. Under it, $\widehat{t}$ has the value $\eta(t)$, which is
positive.
\end{lem}
\begin{proof}
The polynomial $\inner{\widehat{u}_i}{\widehat{u}_i}$ has the value
$\norm{\eta(u_i)}^2\geq0$ under $\eta$, and exactly one nonnegative real has
this square. With these values for $h$, the polynomial $\widehat{t}$ has the
value $\eta(t)$ by the extension of $\eta$ to terms. Here $\eta(t)$ is a
length (Section~\ref{subsec:syntax}).
\end{proof}

\noindent
The sign condition $h_i\geq0$ selects the norm from the two possible square
roots. It permits $h_i=0$; positivity is required of the complete term. Thus,
for $G=I$ and $t=z+\norm{x+c}$, we obtain $\widehat{t}=z+h$ with the
constraint $h\geq0\wedge h^2=(x_1+c_1)^2+\cdots+(x_k+c_k)^2$. Membership in
a window is a polynomial condition: for evaluation tuples $e,e'$ and a
polynomial term $\tau$ we write
\[
W(e',e,\tau):=\inner{e'-e}{e'-e}<\tau^2 ,
\]
which, for a positive value of $\tau$, states that the point denoted by $e'$
lies in the open ball of radius $\tau$ about the point denoted by $e$.

\begin{defi}[Standard translation]\label{def:translation}
Let $\varepsilon$ be a GEL over $\Real^k$. For a formula $\varphi$ of $\ELR$
and an evaluation tuple $e$, the arithmetic formula $T_e(\varphi)$ is defined
by recursion on $\varphi$, after renaming the bound variables of $\varphi$ so
that they do not occur in $e$, and with a fresh $k$-tuple $y$ of variables
and fresh variables $h$ at each bounded modality:
\begin{align*}
T_e(p)&:=\Theta_p(e), & T_e(\varphi_u)&:=T_{e+\widehat{u}}(\varphi),\\
T_e(\neg\varphi)&:=\neg T_e(\varphi), &
T_e(\varphi\circ\psi)&:=T_e(\varphi)\circ T_e(\psi)\quad(\circ\in\{\wedge,\vee\}),\\
T_e(\exists x\,\varphi)&:=\exists x\,T_e(\varphi), &
T_e(\exists z\,\varphi)&:=\exists z\,(z>0\wedge T_e(\varphi)),\\
T_e(\forall x\,\varphi)&:=\forall x\,T_e(\varphi), &
T_e(\forall z\,\varphi)&:=\forall z\,(z>0\rightarrow T_e(\varphi)),
\end{align*}
\begin{align*}
T_e(\mdiamond_t\varphi)&:=\exists h\,\bigl(N_t(h)\wedge
  \exists y\,(W(y,e,\widehat{t})\wedge T_y(\varphi))\bigr),\\
T_e(\boldBox_t\varphi)&:=\exists h\,\bigl(N_t(h)\wedge
  \forall y\,(W(y,e,\widehat{t})\rightarrow T_y(\varphi))\bigr).
\end{align*}
Here $\Theta_p(e)$ results from $\Theta_p$ by substituting the tuple $e$ for
its variables, and a quantifier over a tuple abbreviates quantifiers over its
components. We write $T^{\varepsilon}_e(\varphi)$ when the GEL is to be
named.
\end{defi}

\noindent
A norm constraint is introduced at the bounded modality whose length it
serves. Thus an auxiliary variable for $\norm{x}$ remains inside the
quantifier for $x$. Moving it outside would replace a length that varies
with $x$ by one global choice. In the clause for $\boldBox_t$ the
existential quantifier over $h$ does not relax the condition imposed
throughout the window: by Lemma~\ref{lem:norms} exactly one tuple satisfies
$N_t$, so that $\exists h\,(N_t(h)\wedge\cdots)$ and
$\forall h\,(N_t(h)\rightarrow\cdots)$ are equivalent.

\begin{thm}[Standard translation]\label{thm:translation}
Let $\varepsilon$ be a GEL over $\Real^k$, let $\varphi$ be a formula of
$\ELR$, and let $y$ be a $k$-tuple of arithmetic variables that do not occur
in $\varphi$. For every assignment $\eta$ and every point $a$,
\[
(\varepsilon,\eta,a)\models\varphi
\quad\Longleftrightarrow\quad
\Real,\,\eta[y\mapsto a]\models T_y(\varphi).
\]
The formula $T_y(\varphi)$ is computable from $\varepsilon$ and $\varphi$ in
polynomial time, when the evaluation tuple $e+\widehat{u}$ of a displacement
is represented by a fresh tuple $y'$ of variables with the constraint
$y'=e+\widehat{u}$.
\end{thm}
\begin{proof}
We prove, by induction on $\varphi$, that for every evaluation tuple $e$ and
every assignment $\eta$, extended to the further variables of $e$, the
formula $T_e(\varphi)$ holds in $\Real$ under $\eta$ if and only if
$\varphi$ holds at the point $\eta(e)$ under $\varepsilon$ and $\eta$. The
theorem is the case $e=y$.

For an atomic proposition, $\Theta_p(e)$ holds if and only if
$\eta(e)\in\sem{p}_\varepsilon$, because $\Theta_p$ presents this set. The
Boolean connectives have the same truth conditions in the two languages. For
a displacement $\varphi_u$, the tuple $e+\widehat{u}$ has the value
$\eta(e)+\eta(u)$, and the induction hypothesis applies at this point. For
$\mdiamond_t\varphi$, Lemma~\ref{lem:norms} shows that the only tuple that
satisfies $N_t$ is the tuple of the norms, and that $\widehat{t}$ then has
the value $\eta(t)>0$. For this value, $W(y,e,\widehat{t})$ holds exactly at
the points of the window $B(\eta(e),\eta(t))$. By the induction hypothesis,
$T_y(\varphi)$ holds at such a point exactly when $\varphi$ holds there.
Hence an arithmetic witness for $y$ exists if and only if $\varphi$ holds
somewhere in the window. For $\boldBox_t\varphi$, the same value determines
the same window; the implication imposes the condition $T_y(\varphi)$ at
every point of the window and no condition outside it. A point quantifier
ranges over tuples of coordinates, hence over points. The induction
hypothesis applies to the variant assignments.
The guards $z>0$ restrict the arithmetic quantifiers to the positive reals,
which form the length domain.

Each clause of Definition~\ref{def:translation} adds a bounded number of
symbols, apart from the copy of $\Theta_p$ contributed by each occurrence of
an atomic proposition and the substitution of evaluation tuples. With the
stated representation of the tuples $e+\widehat{u}$, which is equivalent to
the substitution, no term is duplicated, and the translation is computed in
polynomial time. Every evaluation tuple is then fresh, and no bound variable
needs to be renamed.
\end{proof}

\noindent
The proof requires only that $N_t$ determines the norms occurring in $t$
(Lemma~\ref{lem:norms}) and that $W$ defines the window. It therefore
applies to $\EL$ over $\Real^k$ with any norm for which arithmetic formulas
with these two properties are given, such as the polyhedral norms of
Section~\ref{subsec:linear}.

\subsection{Semi-algebraic satisfaction and finite evidence}\label{subsec:evidence}
The translation preserves free variables. It therefore describes the set
of satisfying assignments as well as truth under a given assignment. For a
tuple $\boldsymbol\xi=(\xi_1,\ldots,\xi_m)$ of distinct
variables, let $D_{\boldsymbol\xi}:=\prod_{i=1}^{m}D_i$, where
$D_i=\Real^k$ for a point variable $\xi_i$ and $D_i=\Real_{>0}$ for a length
variable. Then $D_{\boldsymbol\xi}\subseteq\Real^{N}$, where $N$ is $k$
times the number of point variables plus the number of length variables in
$\boldsymbol\xi$. For $\mathbf{a}\in D_{\boldsymbol\xi}$, write
$\varphi[\mathbf{a}]$ for $\varphi$ under an assignment mapping
$\boldsymbol\xi$ to $\mathbf{a}$. Quantification over a tuple abbreviates
successive quantification over its variables.

A \emph{cylindrical algebraic decomposition} (CAD) of $\Real^{N}$ is a finite
partition of $\Real^{N}$ into connected semi-algebraic sets, its
\emph{cells}, such that for every $j<N$ any two cells have equal or disjoint
projections onto the first $j$ coordinates. A CAD is \emph{sign-invariant}
for a finite set of polynomials if each of them has a constant sign,
negative, zero or positive, on every cell. For polynomials with coefficients
in $\Ralg$, a sign-invariant CAD is computable together with a \emph{sample
point} with coordinates in $\Ralg$ in each
cell~\cite{collins1975,bpr2006,basu2017}. Its cells may have any dimension
from $0$ to $N$, and the constructions below retain all of them.

\begin{thm}[Semi-algebraic satisfaction and finite evidence]\label{thm:evidence}
Let $\varepsilon$ be a GEL over $\Real^k$, and let
$\boldsymbol\xi=(\xi_1,\ldots,\xi_m)$ list the free variables of a formula
$\varphi$ of $\ELR$.
\begin{enumerate}[label=\textup{(\roman*)},leftmargin=2.2em]
\item The \emph{satisfaction set} of $\varphi$ in $\varepsilon$,
\[
 S_\varphi:=\{(\mathbf{a},b)\in D_{\boldsymbol\xi}\times\Real^k\mid
                    (\varepsilon,b)\models\varphi[\mathbf{a}]\},
\]
is semi-algebraic, and a presentation of it is computable from $\varepsilon$
and $\varphi$.
\item For every point $a\in\Ralg^k$ one can compute a finite nonempty set
$F\subseteq D_{\boldsymbol\xi}$ of tuples with coordinates in $\Ralg$, called
\emph{finite evidence} at $a$, such that
\begin{align*}
 (\varepsilon,a)\models\exists\boldsymbol\xi\,\varphi
   &\iff (\varepsilon,a)\models\varphi[\mathbf{a}]
                                  \text{ for some }\mathbf{a}\in F,\\
 (\varepsilon,a)\models\forall\boldsymbol\xi\,\varphi
   &\iff (\varepsilon,a)\models\varphi[\mathbf{a}]
                                  \text{ for every }\mathbf{a}\in F.
\end{align*}
The set $F$ may be chosen with one point in each cell contained in
$D_{\boldsymbol\xi}$ of a CAD of $\Real^{N}$ that is sign-invariant for the
polynomials of a presentation of the section of $S_\varphi$ at $a$ and for
the coordinates of the length variables; the truth value of
$\varphi[\mathbf{a}]$ at $a$ is constant on each of these cells.
\end{enumerate}
\end{thm}
\begin{proof}
(i) By Theorem~\ref{thm:translation}, the set $S_\varphi$ is defined by the
conjunction $\Psi$ of $T_y(\varphi)$ with the guards $z>0$ for the length
variables $z$ of $\boldsymbol\xi$, in the variables $y$ and the coordinates
of $\boldsymbol\xi$. Every atomic formula of $\Psi$ is a polynomial equality
or inequality. A coefficient in $\Ralg$ can be replaced by a fresh variable
constrained by its defining polynomial and isolating interval, which
determine it uniquely. This makes $\Psi$ a formula of $\RCF$ with rational
coefficients. Quantifier elimination for real closed fields
produces an equivalent quantifier-free
formula~\cite{tarski1951,seidenberg1954,bpr2006}, and substituting the
algebraic numbers for their variables yields a presentation of $S_\varphi$.
The construction is effective.

(ii) Substitute the coordinates of $a$ for $y$ in a presentation of
$S_\varphi$. The result is a quantifier-free formula $\Psi_a$ in the $N$
coordinates of $\boldsymbol\xi$ such that, for every
$\mathbf{a}\in D_{\boldsymbol\xi}$, $\varphi[\mathbf{a}]$ holds at $a$ if and
only if $\Real\models\Psi_a(\mathbf{a})$. Compute a CAD that is
sign-invariant for the polynomials of $\Psi_a$ and for the coordinates of the
length variables of $\boldsymbol\xi$. The set $D_{\boldsymbol\xi}$ is defined
by the positivity of these coordinates, so that each cell is contained in
$D_{\boldsymbol\xi}$ or disjoint from it. The truth value of the Boolean
combination $\Psi_a$ depends only on the signs of its polynomials and is
therefore constant on each cell. Let $F$ consist of the sample point of each
cell contained in $D_{\boldsymbol\xi}$; these cells partition the nonempty
set $D_{\boldsymbol\xi}$, so that $F$ is nonempty. If some tuple of
$D_{\boldsymbol\xi}$ satisfies $\varphi$ at $a$, then so does the sample
point of its cell, and if some tuple falsifies $\varphi$, then so does the
sample point of its cell. The two equivalences follow. For $m=0$, $F$
consists of the empty tuple.
\end{proof}

\begin{exa}[Translation and evidence on the running example]\label{ex:translation}
For the GEL $\varepsilon_{\ast}$ of Example~\ref{ex:running}, the
presentations are $\Theta_L(y)\equiv(|y|^2\leq0)$,
$\Theta_V(y)\equiv(|y-e_1|^2\leq0)$ and $\Theta_O(y)\equiv(|y|^2\leq4)$,
where $|y|^2$ abbreviates $y_1^2+\cdots+y_k^2$. Consider the formula
$\chi:=\mdiamond\bigl(L\wedge\boldBox_zO\wedge\mdiamond_zV\bigr)$, that is,
$\exists x\,\bigl(L\wedge\boldBox_zO\wedge\mdiamond_zV\bigr)_x$, with the
free length variable $z$. Its parameters contain no norm, so that there are
no norm constraints, and
\[
T_0(\chi)=\exists x\,\Bigl(\Theta_L(x)
\wedge\forall y\,\bigl(W(y,x,z)\rightarrow\Theta_O(y)\bigr)
\wedge\exists y'\,\bigl(W(y',x,z)\wedge\Theta_V(y')\bigr)\Bigr).
\]
The first conjunct fixes $x$ at the origin. The observation condition then
holds precisely for $z\leq2$, the case $z=2$ included because the window is
open, and the proximity condition precisely for $z>1$, so that $\chi$ holds
at the origin exactly for $1<z\leq2$, as in Example~\ref{ex:running}. The
decomposition of the line that is sign-invariant for $z$, $z-1$ and $z-2$
has the cells $(0,1)$, $\{1\}$, $(1,2)$, $\{2\}$ and $(2,\infty)$ inside the
positive half-line, and the sample points $\frac12,1,\frac32,2,3$ form
finite evidence. The formula is false on $\{1\}$ and true on $\{2\}$; the
cells of dimension zero keep these endpoint conditions apart.
\end{exa}

\begin{cor}[Derived atomic propositions]\label{cor:derived}
For every closed formula $\varphi$ of $\ELR$ and every GEL $\varepsilon$ over
$\Real^k$, the support $\sem{\varphi}_\varepsilon$ is effective
semi-algebraic, and a presentation of it is computable. It can therefore be
stored as the single element of a new atomic proposition, and the extended
map is again a GEL.
\end{cor}
\begin{proof}
For a closed formula the tuple $\boldsymbol\xi$ is empty, and the
satisfaction set of Theorem~\ref{thm:evidence}(i) is the support.
\end{proof}

The presentation obtained may leave the class of ball unions: when
$\varepsilon(p)$ consists of the two balls of Example~\ref{ex:coverage}, the
support of $\boldBox_1p$ is the set $\boldBox_1U$ of
Proposition~\ref{prop:closure}.

\begin{algo}[Model checking on a GEL]\label{alg:mc}
\emph{Input:} a GEL $\varepsilon$ over $\Real^k$, a closed formula $\psi$ of
$\ELR$, and a point $a\in\Ralg^k$. \emph{Output:} the truth value of
$(\varepsilon,a)\models\psi$.
\begin{enumerate}[leftmargin=2em]
\item Form $\Theta_p$ for every atomic proposition $p$ that occurs in $\psi$.
\item Compute $T_y(\psi)$, with fresh variables at each bounded modality.
\item Substitute the coordinates of $a$ for $y$.
\item Decide the resulting sentence in the theory of real closed fields, and
return its truth value.
\end{enumerate}
\end{algo}

\noindent
By Theorem~\ref{thm:translation}, correctness of the algorithm reduces to
that of step~4. The arithmetic procedure enters through its specification
alone, as in Theorem~\ref{thm:evidence}, Corollaries~\ref{cor:derived},
\ref{cor:certified} and~\ref{cor:range} and Theorem~\ref{thm:synthesis}.
It is the trusted base of a verdict and may be verified or proof-producing,
as in Coq and HOL~Light for real closed
fields~\cite{cohenmahboubi2012,mclaughlin2005} and in Isabelle/HOL for linear
and quadratic real arithmetic~\cite{nipkow2010,scharager2021}.

Algebraic evidence is necessary in general. In rational time a true
existential specification has a rational witness~\cite{zhang2024temporal}. In space,
rational descriptions can force irrational ones, already for rational
ball-GELs in dimension $k\geq2$ (Example~\ref{ex:irrational-length}).
A general procedure therefore retains the cells of lower dimension and
their algebraic sample points.

Theorem~\ref{thm:evidence}(ii) concerns one block of like quantifiers. It
supplies no independent finite choices for the variables of an alternating
prefix. If $Z$ holds only at the origin, then
$\forall x\,\exists y\,Z_{x+y}$ holds there, with $y$ chosen as $-x$. No
finite list of values for $y$ serves every value of $x$. Evaluation
therefore follows the order of the quantifiers through a single cylindrical
decomposition, taking disjunctions and conjunctions over the fibre cells
above each lower cell.

\subsection{Data complexity and combined complexity}\label{subsec:complexity}
When a specification is applied to a collection of segmentations, the
formula is fixed while the records vary. \emph{Data complexity} measures
this situation: the input consists of the GEL and the point of evaluation.
\emph{Combined complexity} lets the formula vary as well~\cite{vardi1982}.
Rational numbers
are encoded by a numerator and a positive denominator in binary, and
polynomials by their lists of nonzero coefficients.

\begin{thm}[Data complexity]\label{thm:datacomplexity}
Fix the dimension $k$, a rational metric matrix $G$, a closed formula $\psi$
of $\ELR$ with rational constants, and a bound $\Delta$ on the degrees of the
polynomials in presentations. The problem of deciding
$(\varepsilon,a)\models\psi$, for a given rational GEL $\varepsilon$ over
$\Real^k$ with presentations of degree at most $\Delta$ and a given point
$a\in\Rat^k$, is solvable in time polynomial in the binary size of
$\varepsilon$ and $a$. Under the same assumptions, for a fixed formula with
free variables, a presentation of its satisfaction set and finite evidence at
$a$ are computable in polynomial time.
\end{thm}
\begin{proof}
Fix $k$, $G$, $\psi$ and $\Delta$, and let $n$ be the binary size of
$\varepsilon$ and $a$. By Theorem~\ref{thm:translation},
$(\varepsilon,a)\models\psi$ if and only if the sentence $\Omega$ that
results from $T_y(\psi)$ by substituting the coordinates of $a$ for $y$ is
true in $\Real$. We bound the parameters of $\Omega$ that govern the cost of
quantifier elimination.

\emph{Variables.} Let $m_1$ and $m_2$ be the numbers of point and of length
quantifiers of $\psi$, let $\mu$ be its number of bounded modalities, and let
$\nu$ be the number of norms in its parameters. Each point quantifier
contributes $k$ arithmetic variables and each length quantifier one; each
bounded modality contributes the $k$-tuple $y$; and each norm contributes one
variable. Hence $\Omega$ has
\[
 M:=k(m_1+\mu)+m_2+\nu
\]
variables in the substitution form of Definition~\ref{def:translation}, used
in this proof, and $k$ more for each displacement of $\psi$ with the fresh
tuples of Theorem~\ref{thm:translation}; both depend on $k$ and $\psi$ only.

\emph{Polynomials.} Each occurrence of an atomic proposition $p$ in $\psi$
contributes one copy of $\Theta_p$, in which the evaluation tuple accumulated
along its branch is substituted for the variables; every other clause of
Definition~\ref{def:translation} contributes a number of polynomials that is
bounded in terms of $\psi$. Substituting polynomials of degree at most $1$
does not raise the degree.
The polynomials of $\Omega$ thus have degree at most $\max\{\Delta,2\}$,
where $2$ accounts for the windows and norm constraints. A polynomial of
bounded degree in $M$
variables has a bounded number of monomials. Each substituted polynomial is
therefore computed in polynomial time and has coefficients of bit length
polynomial in $n$. Altogether $\Omega$ contains $O(n)$ polynomials, and
clearing denominators yields integer coefficients of bit length polynomial
in $n$.

\emph{Elimination.} Bring $\Omega$ into prenex form; this changes neither its
polynomials nor its variables. For a prenex sentence with $M$ variables and
$g$ polynomials of degree at most $\delta$ with integer coefficients of bit
length at most $\beta$, in which atomic formulas occur $q$ times, quantifier
elimination by cylindrical algebraic decomposition runs in time
\[
 O\bigl((2\delta)^{2^{2M+8}}\,g^{2^{M+6}}\,\beta^{3}\,q\bigr)
\]
\cite{collins1975}; see~\cite{bpr2006,basu2017} for sharper bounds. With $M$
and $\delta$ fixed and $g$, $\beta$ and $q$ polynomial in $n$, this bound is
polynomial in $n$. The procedure computes a sign-invariant CAD of $\Real^{M}$
for the polynomials of $\Omega$, evaluates the quantifier-free matrix at the
sample points, and propagates truth values from the innermost coordinates
outwards, as described in Section~\ref{subsec:evidence}.

For a fixed formula with free variables, order the coordinates of the
decomposition so that the free arithmetic variables come first. Their number
is fixed, so that the same bound applies. The cells of the induced
decomposition of the parameter space, labelled with truth values, describe
the satisfaction set. The elimination procedure converts them into a
quantifier-free formula; sample points are represented by isolating
intervals or Thom encodings~\cite{collins1975,bpr2006}. For finite
evidence, retain the sample points of all cells contained in
$D_{\boldsymbol\xi}$; the cells on which the formula is false are needed
for the universal equivalence of Theorem~\ref{thm:evidence}(ii).
\end{proof}

For rational ball-GELs the presentations have degree $2$, and the theorem
applies with $\Delta=2$. For general presentations with binary exponents,
the degree bound is a substantive restriction. The exponent of the
polynomial bound depends on the formula and on the dimension: the stenosis
formula of Section~\ref{subsec:library}, with three point quantifiers, one
length quantifier and two bounded modalities, has $M=5k+1$. The theorem
parallels the result for QEL~\cite{qel}. With a fixed number of
variables, the theory of real closed fields is also decidable in the
parallel class NC~\cite{benor1986}.

\begin{prop}[Combined complexity]\label{prop:combined}
In every dimension $k\geq1$, the combined complexity of model checking for
$\ELR$ is PSPACE-hard. This holds already for one fixed GEL with a single
atomic proposition, at the origin, and for formulas without bounded
modalities. For rational GELs with exponents written in unary, rational
constants and points of evaluation, and a rational metric matrix, the
problem is in EXPSPACE, uniformly in $k$.
\end{prop}
\begin{proof}
\emph{Lower bound.} We reduce from the validity problem for quantified
Boolean formulas, which is PSPACE-complete~\cite{stockmeyer1973}. Let the
atomic proposition $Z$ hold at the origin only, and put
\[
 \Bool(x):=Z_{x}\vee Z_{x+e_1}.
\]
At the origin this formula holds exactly under the assignments that map $x$
to $0$ or to $-e_1$, which encode true and false, respectively. Translate a
quantified Boolean formula by assigning a point variable $x_b$ to each
Boolean variable $b$, replacing each occurrence of $b$ by $Z_{x_b}$,
retaining the Boolean connectives, and translating the quantifiers by
\[
 (\exists b\,\chi)^{*}:=\exists x_b\,(\Bool(x_b)\wedge\chi^{*})
 \qquad\text{and}\qquad
 (\forall b\,\chi)^{*}:=\forall x_b\,(\Bool(x_b)\rightarrow\chi^{*}).
\]
Quantifiers do not move the point of evaluation, so that every formula
$Z_{x_b}$ is evaluated at the origin, where it holds if and only if $x_b$
denotes the encoding of true. By induction on $\chi$, for every assignment
that maps each free $x_b$ to one of the two encoding points, $\chi^{*}$ holds
at the origin if and only if $\chi$ holds under the encoded Boolean
assignment. The Boolean cases are immediate. For an existential quantifier,
the guard admits precisely the two encoding points; for a universal
quantifier, every other point satisfies the implication vacuously, and the
same two choices remain to be checked. Each constructor adds a bounded number
of symbols, so that the reduction is computable in linear time. The argument
does not use the metric, so the bound holds for arbitrary $k$ and $G$.

\emph{Upper bound.} The sentence that results from $T_y(\psi)$ by
substituting the coordinates of the point of evaluation is computable in
polynomial time (Theorem~\ref{thm:translation}); the dimension $k$ is at most
the size of the input, since the point of evaluation has $k$ coordinates. The
theory of real closed fields is decidable in exponential
space~\cite{benor1986}.
\end{proof}

The exact combined complexity remains open between these bounds; in the
discrete logic over the natural numbers, model checking over finite words is
PSPACE-complete in combined complexity~\cite{DZ26}.

\subsection{The line and polyhedral norms: evaluation in linear
arithmetic}\label{subsec:linear}
Euclidean balls in higher dimensions call for nonlinear arithmetic. When the
balls of a norm are polytopes, its graph and windows have linear
descriptions. Evaluation then returns to the arithmetic used for QEL. Let
$\mathcal{F}$ be a finite set of linear forms on $\Real^k$ with rational
coefficients that is closed under negation and spans the dual space of
$\Real^k$. Then
$\norm{v}_{\mathcal{F}}:=\max_{f\in\mathcal{F}}f(v)$
is a norm, a \emph{rational polyhedral norm}, and $\Real^k$ with the metric
$d_{\mathcal{F}}(a,b):=\norm{a-b}_{\mathcal{F}}$ and the length domain
$\Real_{>0}$ is a metric monoid, which we denote by
$\Real^k_{\mathcal{F}}$. An example is the maximum norm, with the $2k$
forms $\pm v_i$. A norm on the line has this form:
$\gamma|v|$ is the maximum of the two forms $\gamma v$ and $-\gamma v$, and it
is a rational polyhedral norm when $\gamma$ is rational. In the logic
$\EL(\Real^k_{\mathcal{F}})$, in full mode and with rational constants, the
window of length $\ell$ at $a$ is the open polytope
$\{b\mid\norm{b-a}_{\mathcal{F}}<\ell\}$. A subset of $\Real^k$ is
\emph{rational semilinear} if it is presented by a Boolean combination of
affine equalities and inequalities with rational coefficients, and a GEL is
rational semilinear if all its elements are.

\begin{prop}[Linear arithmetic and rational evidence]\label{prop:linear}
Let $\varepsilon$ be a rational semilinear GEL over $\Real^k$, interpreted
in the logic $\EL(\Real^k_{\mathcal{F}})$.
\begin{enumerate}[label=\textup{(\roman*)},leftmargin=2.2em]
\item Every satisfaction set is rational semilinear and effectively
computable.
\item At every point of $\Rat^k$, the finite evidence of
Theorem~\ref{thm:evidence}\textup{(ii)} can be chosen in $\Rat^{N}$.
\item Under rational assignments and at rational points, a formula holds
over $\Real^k$ if and only if it holds over the metric monoid $\Rat^k$ with
the same norm and the length domain $\Rat_{>0}$.
\end{enumerate}
In particular, the three statements hold for the Euclidean line, that is,
for $\EL(\Real)$ with a rational $\gamma$ and for GELs whose elements are
intervals with rational endpoints.
\end{prop}
\begin{proof}
The graph of the norm is defined by the linear formula
$\bigwedge_{f\in\mathcal{F}}h\geq f(v)\wedge\bigvee_{f\in\mathcal{F}}h=f(v)$,
which determines $h=\norm{v}_{\mathcal{F}}$ uniquely, and membership of $y'$
in the window of length $\tau$ at $y$ is the conjunction of the conditions
$f(y'-y)<\tau$ for $f\in\mathcal{F}$. With these formulas in place of the
quadratic equations of the norm constraints and of the conditions
$W(y',y,\tau)$, the translation of Definition~\ref{def:translation} consists
of linear constraints with rational coefficients, and the proof of
Theorem~\ref{thm:translation} goes through without change. Quantifier
elimination for ordered divisible abelian groups produces an equivalent
Boolean combination of rational affine
conditions~\cite{ferrante1975,loos1993,marker2002}; this is
the elimination used for QEL~\cite{qel}. This proves (i).

For (ii), the hyperplanes of these affine conditions, together with the
hyperplanes $z=0$ of the length variables, partition the space of parameters
into finitely many relatively open polyhedral faces of all dimensions, on
each of which the truth value is constant. Each face contains a rational
point: its equalities have a rational solution and a rational basis of their
solution space, by Gaussian elimination; rational combinations of this basis
are dense in the real affine solution space; and the strict inequalities of
the face define a nonempty relatively open subset of that space. One rational
point from each face contained in $D_{\boldsymbol\xi}$ forms finite evidence,
by the argument of Theorem~\ref{thm:evidence}.

For (iii), $\norm{\cdot}_{\mathcal{F}}$ maps $\Rat^k$ to $\Rat$, so terms
denote rational points and lengths under rational assignments. The logic
over $\Rat^k$ is therefore well defined. The proof of
Theorem~\ref{thm:translation} applies with $\Rat$ in place of $\Real$.
Linear quantifier elimination is valid in both ordered groups. After
rational values are substituted for the free variables, the resulting
quantifier-free rational conditions have the same truth values in $\Rat$ and
in $\Real$. The last assertion is the case of the two forms $\gamma v$ and
$-\gamma v$.
\end{proof}

For the Euclidean norm in dimension $k\geq2$, neither linearity nor
rational evidence survives, already for the simplest rational input.

\begin{exa}[An irrational length from one rational point]\label{ex:irrational-length}
Let $k\geq2$ and $G=I$, and let $p$ hold at the point $e_1+e_2$ only; this
interpretation is presented by a rational ball-GEL with a single element of
radius $0$. For length variables $z,w$ put
\[
\psi:=\boldBox_z\neg p\wedge\forall w\,\neg\boldBox_{z+w}\neg p .
\]
At the origin, $\boldBox_z\neg p$ holds exactly when the window of length $z$
omits the point $e_1+e_2$, that is, when $z\leq\sqrt2$, and the second
conjunct holds exactly when $z+w>\sqrt2$ for every length $w$, that is, when
$z\geq\sqrt2$. Hence $\exists z\,\psi$ holds at the origin, its only witness
is the length $\sqrt2$, and every finite evidence contains this irrational
number.
\end{exa}

Rational polyhedral supports and windows, among them rational interval
records on the line, thus admit linear evaluation with rational evidence,
whereas Euclidean balls in dimension $k\geq2$ require the algebraic
treatment of Sections~\ref{subsec:translation}--\ref{subsec:complexity}.

\section{Enclosures: Certified Evaluation on Approximate Records}\label{sec:enclosures}

Exact model checking settles truth in the given geometry. What this says
about an image depends on the relation to the source annotation. A compact
ball representation alone does not certify that relation. Lower and upper
bounds on the atomic supports carry uncertainty into evaluation. By
Theorem~\ref{thm:enclosure} these bounds propagate through formulas, allowing
the checker to certify conclusions about an interpretation known only
within an enclosure.

\subsection{A finite image as an exact reference geometry}\label{subsec:cells}
Let a finite image have samples with indices $j\in J$, coordinates in a
calibrated frame, and an assigned cell $Q_j\subseteq\Real^k$ for each sample.
The \emph{cell reconstruction} of the image is the interpretation in which a
label $p$ of the image, read as an atomic proposition, holds exactly on the
union of the cells $Q_j$ of the samples to which $p$ is assigned. Rational
boxes, with any choice of half-open faces, have rational presentations, so
that the cell reconstruction is presented by a rational GEL, and
Section~\ref{sec:modelchecking} applies to it directly. The cells may be
made pairwise disjoint by a fixed convention for the ownership of
boundaries, or they may be closed and overlap on shared faces. The evaluator
and its reference use the same convention.

A finite model on the samples alone is a different object. Let $G=I$, let
$X:=\{-1,0,1\}^k$ be a sample set, and let $p$ hold only at the origin. In
the finite model with domain $X$, the window of length $1/2$ at the origin
contains no other sample, so that a universal check returns true; in
$\Real^k$, $\boldBox_{1/2}\,p$ fails there, since $\frac14e_1$ lies in the
window and does not satisfy $p$. Both answers are correct for their
semantics. Inferring continuous coverage from sampled data requires a
specified reconstruction or an enclosure.

\subsection{Lower and upper GELs}\label{subsec:enclosure}

\begin{defi}[Enclosure]\label{def:enclosure}
An \emph{enclosure} of an interpretation $\alpha$ is a pair
$(\varepsilon^{-},\varepsilon^{+})$ of GELs over $\Real^k$ such that
\[
\sem{p}_{\varepsilon^{-}}\subseteq\sem{p}_{\alpha}\subseteq\sem{p}_{\varepsilon^{+}}
\qquad\text{for every }p\in\Prop.
\]
The GELs $\varepsilon^{-}$ and $\varepsilon^{+}$ are the \emph{lower} and the
\emph{upper bound} of the enclosure, and $\alpha$ is \emph{enclosed} by it.
\end{defi}

\noindent
The enclosed interpretation $\alpha$ need not have a finite presentation.
An enclosure may certify the approximation of a cell reconstruction, express
uncertainty about an annotation, or describe unobserved space
conservatively.

Let $\varphi$ be a formula, written without abbreviations. An occurrence of
an atomic proposition in $\varphi$ is \emph{positive} if it lies in the
scope of an even number of negations, and \emph{negative} otherwise. Let
$\Prop^{\pm}:=\{p^{-},p^{+}\mid p\in\Prop\}$ be a set of new atomic
propositions, and let $\varepsilon^{\pm}$ be the GEL over $\Prop^{\pm}$ with
$\varepsilon^{\pm}(p^{-}):=\varepsilon^{-}(p)$ and
$\varepsilon^{\pm}(p^{+}):=\varepsilon^{+}(p)$. Let $\varphi^{-}$ result from
$\varphi$ by replacing each positive occurrence of an atomic proposition $p$
by $p^{-}$ and each negative occurrence by $p^{+}$, and let
$\varphi^{+}$ be defined with the roles of $p^{-}$ and $p^{+}$ exchanged, so
that $(\neg\varphi)^{-}=\neg(\varphi^{+})$ and
$(\neg\varphi)^{+}=\neg(\varphi^{-})$. The
\emph{lower} and the \emph{upper evaluation} of $\varphi$ are the supports
\[
\sem{\varphi}^{-}_{\eta}:=\sem{\varphi^{-}}_{\varepsilon^{\pm},\eta}
\qquad\text{and}\qquad
\sem{\varphi}^{+}_{\eta}:=\sem{\varphi^{+}}_{\varepsilon^{\pm},\eta}.
\]
Both are ordinary supports in an auxiliary GEL, computed by the checker of
Section~\ref{sec:modelchecking}. The semantics remains two-valued;
uncertainty enters through the bounds on the interpretation.

\begin{thm}[Propagation of bounds]\label{thm:enclosure}
For every enclosure of an interpretation $\alpha$, every formula $\varphi$
and every assignment $\eta$,
\[
\sem{\varphi}^{-}_{\eta}\subseteq\sem{\varphi}_{\alpha,\eta}
\subseteq\sem{\varphi}^{+}_{\eta}.
\]
The formula may contain negations at any depth and quantifiers of both sorts
in arbitrary alternation.
Moreover, the bounds are monotone in the enclosure: if the lower bound of a
second enclosure has larger supports and its upper bound has smaller
supports, then its lower evaluation of $\varphi$ contains
$\sem{\varphi}^{-}_{\eta}$ and its upper evaluation is contained in
$\sem{\varphi}^{+}_{\eta}$.
\end{thm}
\begin{proof}
By induction on $\varphi$, using the identities of
Proposition~\ref{prop:semantics}; the two inclusions are proved together.
For an atomic proposition $p$ the three sets are
$\sem{p}_{\varepsilon^{-}}$, $\sem{p}_\alpha$ and
$\sem{p}_{\varepsilon^{+}}$. For a negation $\neg\psi$ they are the
complements of $\sem{\psi}^{+}_{\eta}$, $\sem{\psi}_{\alpha,\eta}$ and
$\sem{\psi}^{-}_{\eta}$, and complementation reverses the two inclusions
that the induction hypothesis gives for $\psi$. A
displacement translates the three sets by the same point. Intersection and
union are monotone in each argument. Dilation and erosion are monotone by
Proposition~\ref{prop:morph}(i): a witness in a window remains available when
the support grows, and a window contained in a support remains contained in
a larger one. Hence the inclusions pass through $\wedge$, $\vee$,
$\mdiamond_t$ and $\boldBox_t$, the length $\eta(t)$ being the same in the
three evaluations. The support of $\exists x\,\psi$ is the union of the
supports of $\psi$ over all points for $x$, and the support of
$\forall x\,\psi$ is their intersection; the same holds for the length
quantifiers. The induction hypothesis gives the inclusions for $\psi$ under
every variant assignment, and unions and intersections of arbitrary families
preserve inclusions. No finiteness of the domain of quantification is used.
The same argument, started from the inclusions between the two enclosures at
the atomic propositions, proves the last assertion.
\end{proof}

\begin{cor}[Certified verdicts]\label{cor:certified}
Let $\varphi$ be a closed formula, and let $a\in\Ralg^k$. If
$a\in\sem{\varphi}^{-}$, then $(\alpha,a)\models\varphi$ for every
interpretation $\alpha$ that is enclosed by
$(\varepsilon^{-},\varepsilon^{+})$. If $a\notin\sem{\varphi}^{+}$, then
$(\alpha,a)\not\models\varphi$ for every such interpretation. Both
conditions are decidable.
\end{cor}
\begin{proof}
The first two assertions restate Theorem~\ref{thm:enclosure}. The two
conditions are instances of the model-checking problem for the GEL
$\varepsilon^{\pm}$, which is decidable by Theorem~\ref{thm:translation}.
\end{proof}

Formulas without negation are monotone in the atomic
supports~\cite{qel}. The polarity of occurrences extends the argument to
arbitrary formulas; ADNF (Proposition~\ref{prop:laws}(v)) merely displays it at the
literals. A report based on the
corollary has three outcomes: certified truth throughout the enclosure,
certified falsity throughout it, and an inconclusive result. The last is
not a third truth value of $\EL$. Negative conditions in unobserved regions
are a
typical use: if the upper bound of $V$ includes the unobserved locations at
which a vessel has not been excluded, the lower evaluation of $\neg V$
omits that space. Missing information is then not converted into a claim of
absence. The compositional test may remain inconclusive even when every
enclosed interpretation agrees on the query.

For instance, let $k=2$ and $G=I$, let $p$ hold exactly on the square
$[-1,1]^2$, and take the bounds $\varepsilon^{-}(p)=(D(0,1))$ and
$\varepsilon^{+}(p)=(D(0,\sqrt2))$, the upper one with the rational
presentation $y_1^2+y_2^2\leq2$. At the origin $\boldBox_1p$ is certified
true, since $\boldBox_1D(0,1)=\{0\}$, and at $2e_1$ the formula $p$ is
certified false. At the corner $e_1+e_2$, which lies in the upper bound and
outside the lower one, the enclosure decides neither $p$ nor the valid
formula $p\vee\neg p$, whose lower evaluation
$D(0,1)\cup D(0,\sqrt2)^{c}$ omits the corner.

\subsection{Certifying the construction and the selection of balls}\label{subsec:cover}
For an effective semi-algebraic target $S$ and closed balls $E^{-}_i$ and
$E^{+}_j$, the two claims
$\bigcup_iE^{-}_i\subseteq S$ and $S\subseteq\bigcup_jE^{+}_j$
are sentences of the theory of real closed fields. Quantifier elimination
decides them and yields a counterexample when an inclusion fails
(Theorem~\ref{thm:evidence}). These are the obligations of a geometric
compression algorithm, whatever heuristics propose its balls. Testing at
cell centres does not discharge them. The selection of
balls can be organized as a finite set-cover problem. Let
$R_1,\ldots,R_n$ be pieces of the target and, for each candidate ball
$E_i$, let $J_i:=\{j\mid R_j\subseteq E_i\}$, with each inclusion decided
exactly. The greedy set-cover algorithm then uses at most $H_m$ times as
many balls as a smallest selection whose sets $J_i$ cover
$\{1,\ldots,n\}$, where $m:=\max_i|J_i|$ and
$H_m:=\sum_{j=1}^{m}1/j$~\cite{chvatal1979}. The guarantee is relative to
the specified pieces, since several balls may cover a piece jointly
(Example~\ref{ex:coverage}).

\subsection{Enclosures from a tolerance}\label{subsec:tolerance}
Geometric approximations often come with a tolerance in place of explicit
inclusions. Given an approximation $\widehat{S}$ to a support $S$ and a
tolerance $\tau>0$, one might take the erosion $\boldBox_\tau\widehat{S}$
and the dilation $\mdiamond_\tau\widehat{S}$ as lower and upper bounds. The
following characterization identifies the distance conditions that this
choice requires.

\begin{prop}[Enclosures from a tolerance]\label{prop:tolerance}
Let $S,\widehat{S}\subseteq\Real^k$ and $\tau>0$. Then
\begin{align*}
S\subseteq\mdiamond_\tau\widehat{S}
 &\iff\dist(b,\widehat{S})<\tau\ \text{ for all }b\in S,\\
\boldBox_\tau\widehat{S}\subseteq S
 &\iff\dist(b,\widehat{S}^{c})<\tau\ \text{ for all }b\in S^{c}.
\end{align*}
\end{prop}
\begin{proof}
The first equivalence is Proposition~\ref{prop:morph}(vi), applied to
$\widehat{S}$. For the second, $\boldBox_\tau\widehat{S}\subseteq S$ if and
only if $S^{c}\subseteq(\boldBox_\tau\widehat{S})^{c}
=\mdiamond_\tau(\widehat{S}^{c})$, by duality, and
Proposition~\ref{prop:morph}(vi) applies to $\widehat{S}^{c}$.
\end{proof}

\noindent
The upper bound concerns the one-sided distance from $S$ to $\widehat{S}$;
the lower bound concerns the distance between the complements. Hausdorff
control of $S$ and $\widehat{S}$ does not supply that bound. A small cavity shows the
difference. Let $G=I$, $\widehat{S}:=D(0,2)$ and, for $0<\tau<1$,
$S_\tau:=D(0,2)\setminus B(0,\tau)$. The Hausdorff distance between $S_\tau$
and $\widehat{S}$ is $\tau$. Yet $\boldBox_1p$ holds at the origin if $p$
holds exactly on $\widehat{S}$ and fails if $p$ holds exactly on $S_\tau$,
however small $\tau$ is. No erosion
$\boldBox_{\tau'}\widehat{S}=D(0,2-\tau')$ with $\tau'\leq2$ is a lower
bound for $S_\tau$. This is why Theorem~\ref{thm:enclosure} assumes
inclusions.

\section{Bounded Synthesis and the Scope of Decidability}\label{sec:synthesis}

In model checking, the atomic interpretations are given; in synthesis,
they are unknown. Theorem~\ref{thm:translation} does not by itself decide
satisfiability. With a fixed budget of balls, however, the unknown geometry
has finitely many real parameters, and the same arithmetic methods apply.
Section~\ref{subsec:scope} shows that the budget cannot simply be dropped.

\subsection{Synthesis with a prescribed budget of balls}\label{subsec:budget}
For each atomic proposition $p$, fix a \emph{budget} $n_p\geq0$. Introduce
unknown centres $c_{p,i}\in\Real^k$, radii $\rho_{p,i}$ and activation values
$\omega_{p,i}$ for $1\leq i\leq n_p$, subject to
$\rho_{p,i}\geq0$ and $\omega_{p,i}(\omega_{p,i}-1)=0$.
The activation value $0$ disables an element, so that an atomic proposition
may receive fewer than $n_p$ balls, or none. The atomic proposition $p$ is
presented by
\begin{equation}\label{eq:synthesis}
\Theta_p(y):=\bigvee_{i=1}^{n_p}\bigl(\omega_{p,i}=1\wedge
\inner{y-c_{p,i}}{y-c_{p,i}}\leq\rho_{p,i}^2\bigr),
\end{equation}
where the empty disjunction is false. These unknowns parameterize one GEL:
occurrences of the same atomic proposition share the parameters.

\begin{thm}[Bounded synthesis of ball-GELs]\label{thm:synthesis}
The following problem is decidable. Given a closed formula $\psi$ of $\ELR$,
budgets $(n_p)_{p\in\Prop}$, a point $a\in\Ralg^k$, and an arithmetic formula
$\Xi$ with coefficients in $\Ralg$ over the parameters of the GEL, decide
whether there is a ball-GEL $\varepsilon$ over $\Real^k$ with at most $n_p$
balls for each atomic proposition $p$, whose parameters satisfy $\Xi$, such
that $(\varepsilon,a)\models\psi$. If such a ball-GEL exists, one with
centres and radii in $\Ralg$ can be computed.
\end{thm}
\begin{proof}
Let $\boldsymbol\theta$ be the tuple of all parameters $c_{p,i}$,
$\rho_{p,i}$ and $\omega_{p,i}$. With the presentations~\eqref{eq:synthesis},
Definition~\ref{def:translation} yields a formula $T_y(\psi)$ in which the
parameters are free. The proof of Theorem~\ref{thm:translation} applies for
each real value of $\boldsymbol\theta$, since the truth of an atomic
proposition at a point is the displayed polynomial condition. The synthesis
problem is therefore the truth in $\Real$ of the sentence
\[
\exists\boldsymbol\theta\,\Bigl(\bigwedge_{p,i}
 \bigl(\rho_{p,i}\geq0\wedge\omega_{p,i}(\omega_{p,i}-1)=0\bigr)
 \wedge\Xi(\boldsymbol\theta)\wedge T_a(\psi)\Bigr),
\]
which is decided by quantifier elimination. The quantifier over
$\boldsymbol\theta$ stands outside the entire translation, so that one GEL,
and not a different GEL for each inner assignment, satisfies the formula.

If the sentence holds, the set of admissible tuples of parameters is a
nonempty semi-algebraic set with a computable presentation, and a
sign-invariant CAD for the polynomials of this presentation supplies a sample
point in it with coordinates in $\Ralg$. The elements with
$\omega_{p,i}=1$ at this point form a ball-GEL that satisfies $\psi$ at $a$.
\end{proof}

The procedure is uniform in the budgets, which are part of the input. The
theorem offers no small complexity bound as they grow. Rational parameters
need not suffice. Let $k\geq2$, $G=I$ and
$\Prop=\{p,p'\}$, each with the budget $1$; let $\Xi$ fix the ball of $p$ to
be the point $D(e_1+e_2,0)$ and the centre of the ball of $p'$ to be the
origin, and let $\psi:=\mdiamond(p\wedge\partial p')$, with the boundary
operator of Section~\ref{subsec:topology}. The formula states that the point
$e_1+e_2$ lies on the boundary of the ball $D(0,\rho)$. The only solution
has the irrational radius $\rho=\sqrt2$, although the given data are
rational. To require rational parameters is therefore to pose a different
problem, which real quantifier elimination does not settle.

\subsection{Scope of the decidability results}\label{subsec:scope}
The decidability results rely either on a specified interpretation
(Theorem~\ref{thm:translation}) or on a specified family with finitely many
real parameters (Theorem~\ref{thm:synthesis}). To see what happens without
these restrictions, we distinguish unrestricted satisfiability from
satisfiability over finite geometric presentations. The former is the notion
of Definition~\ref{def:entail} and admits all interpretations. A formula is
\emph{satisfiable over GELs} if it holds at some point under some GEL and
some assignment. It is \emph{satisfiable over ball-GELs} if that GEL can be
chosen to consist of balls.

\begin{prop}[Limits of decidability]\label{prop:limits}
Let $k\geq1$.
\begin{enumerate}[label=\textup{(\roman*)},leftmargin=2.2em]
\item For an infinite set $\Prop$, satisfiability for $\ELR$ is
$\Pi^0_1$-hard, already for closed formulas with one quantifier and no
bounded modality. Hence it is undecidable and not recursively enumerable.
\item For a finite set $\Prop$ with at least two elements, satisfiability
over GELs is $\Sigma^0_1$-complete, and so is satisfiability over ball-GELs.
\end{enumerate}
\end{prop}
\begin{proof}
(i) We express the tiling problem as in~\cite{zhang2024temporal}, with a guard that marks a
copy of $\Nat^2$. Let $\Gamma\subseteq\Prop$ be a finite set of \emph{tiles}
with relations $\mathcal{H},\mathcal{V}\subseteq\Gamma\times\Gamma$ of
admissible horizontal and vertical pairs. A \emph{tiling} of the quadrant is a
map $\iota\colon\Nat^2\to\Gamma$ with
$(\iota(i,j),\iota(i+1,j))\in\mathcal{H}$ and
$(\iota(i,j),\iota(i,j+1))\in\mathcal{V}$ for all $i,j$. Whether one exists
is $\Pi^0_1$-complete, as for tilings of the plane~\cite{berger1966}. Let
$c_1,c_2$ be point constants such that the points $ic_1+jc_2$ with
$i,j\in\Nat$ are pairwise distinct, such as $e_1$ and $\sqrt2\,e_1$, as
$\sqrt2$ is irrational. Let $Y\in\Prop\setminus\Gamma$, and put
\[
\Phi_\Gamma:=Y\wedge\boldBox\Bigl(Y\rightarrow\Bigl(Y_{c_1}\wedge Y_{c_2}\wedge
 \bigwedge_{p\neq p'}\neg(p\wedge p')\wedge
 \bigvee\bigl\{\,p\wedge p'_{c_1}\wedge p''_{c_2}\bigm|
 (p,p')\in\mathcal{H},\ (p,p'')\in\mathcal{V}\,\bigr\}\Bigr)\Bigr),
\]
where $p,p',p''$ range over $\Gamma$. If $\Phi_\Gamma$ holds at a point $a$,
then $Y$ holds at all points $a_{ij}:=a+ic_1+jc_2$ with $i,j\in\Nat$, exactly
one tile holds at each of them, and the map $\iota$ that assigns this tile
to $(i,j)$ is a tiling: the disjunct that holds at $a_{ij}$ puts
$(\iota(i,j),\iota(i+1,j))$ in $\mathcal{H}$ and
$(\iota(i,j),\iota(i,j+1))$ in $\mathcal{V}$. Conversely, for a tiling
$\iota$ let $Y$ hold exactly at the pairwise distinct points $ic_1+jc_2$,
and let $\iota(i,j)$ be the only tile that holds there; then
$\Phi_\Gamma$ holds at the origin.

(ii) A formula $\varphi$ with the free variables $\boldsymbol\xi$ is
satisfiable over GELs if and only if the closed formula
$\mdiamond\exists\boldsymbol\xi\,\varphi$ holds at the origin under some GEL
(Lemma~\ref{lem:unbounded}). The GELs can be enumerated, as can the
ball-GELs, and model checking is decidable (Theorem~\ref{thm:translation});
hence both problems are in $\Sigma^0_1$.

For the hardness we reduce the finite satisfiability of first-order sentences
without equality over one binary relation symbol $\mathsf{E}$, which is
$\Sigma^0_1$-complete by Trakhtenbrot's
theorem~\cite{trakhtenbrot1950,kirst2022}. Let $p,p'\in\Prop$ be distinct,
and put
\[
\zeta(y):=\mdiamond\bigl(\mdiamond_{1+\norm{y}}\,p\wedge\neg\mdiamond_1p\bigr),
\qquad
\Lambda:=\mdiamond p\wedge\mdiamond\neg\mdiamond_1p\wedge
\forall y\,\Bigl(\zeta(y)\rightarrow
\boldBox\bigl((p\wedge p_{2y})\rightarrow(\neg\mdiamond_1p)_y\bigr)\Bigr).
\]
Let $\Lambda$ hold under an interpretation $\alpha$, and let
$S:=\sem{p}_\alpha$ and $U:=\sem{\mdiamond_1p}_\alpha$. By the first two
conjuncts, there are points $q_0\in S$ and $q_1\notin U$. We claim that
$\zeta(y)$ holds under an assignment $\eta$
if and only if $\eta(y)\neq0$. If $\eta(y)=0$, the two windows of $\zeta(y)$
coincide. Otherwise, the distance from $S$ is continuous and passes from $0$
at $q_0$ to at least $1$ at $q_1$; by the intermediate value theorem the
segment from $q_0$ to $q_1$ contains a point
$q$ with $\dist(q,S)=1<1+\norm{\eta(y)}$, so that
$\mdiamond_{1+\norm{y}}\,p\wedge\neg\mdiamond_1p$ holds at $q$. By the third
conjunct, the midpoint $a+b$ of two distinct points $a$ and $a+2b$ of $S$
therefore has a distance of at least $1$ from $S$, whence $\norm{b}\geq1$:
distinct points of $S$ are at a distance of at least $2$. If $\alpha$ is
presented by a GEL, then $S$ is semi-algebraic and has finitely many
connected components~\cite{bpr2006}; being discrete, it is finite.

For a first-order formula $\vartheta$, whose variables we read as point
variables, let $\vartheta^{\circ}$ result from $\vartheta$ by replacing every
atomic formula $\mathsf{E}(x,y)$ by $p'_{2x+y}$, every quantifier
$\exists x\,\vartheta'$ by $\exists x\,(p_x\wedge\vartheta'^{\circ})$, and
every quantifier $\forall x\,\vartheta'$ by
$\forall x\,(p_x\rightarrow\vartheta'^{\circ})$. Let $\vartheta$ be a
sentence. If $\Lambda\wedge\vartheta^{\circ}$ holds at a point $a$ under a
GEL $\varepsilon$, then the set $S-a$ is finite and nonempty, and with the
relation $\{(b,b')\in(S-a)^2\mid a+2b+b'\in\sem{p'}_\varepsilon\}$ it is a
model of $\vartheta$: by induction, a subformula $\vartheta'$ holds in this
structure under an assignment $\eta$ with values in $S-a$ if and only if
$(\varepsilon,\eta,a)\models\vartheta'^{\circ}$. Conversely, let $\vartheta$
have a model with the domain $\{1,\ldots,n\}$ and the relation $\varrho$.
Let $c:=e_1/\norm{e_1}$, a point of $\Ralg^k$ of norm $1$, and
$a_i:=4^ic$. Let $\varepsilon(p)$ list the balls $D(a_i,0)$ for
$1\leq i\leq n$, let $\varepsilon(p')$ list the balls $D(2a_i+a_j,0)$ for
$(i,j)\in\varrho$, and let the other atomic propositions have the empty
list. The support of $p$ is nonempty and bounded. The midpoint of $a_i$
and $a_j$, $i<j$, lies between $a_{j-1}$ and $a_j$ at a distance of at least
$6$ from both, so $\Lambda$ holds. The map
$(i,j)\mapsto2\cdot4^i+4^j$ is injective, so $\vartheta^{\circ}$ holds at
the origin by the same induction. Since $\varepsilon$ is a ball-GEL,
$\vartheta$ has a finite model if and only if $\Lambda\wedge\vartheta^{\circ}$
is satisfiable over GELs, and if and only if it is satisfiable over
ball-GELs.
\end{proof}

\noindent
Part~(ii) makes the role of the budget precise: no computable function of the
formula bounds the ball budgets that suffice in Theorem~\ref{thm:synthesis}.
It also implies that the formulas valid at every point of every GEL are not
recursively enumerable. Thus no proof system with a decidable notion of
proof can be both sound and complete for this notion of validity. The same
holds over LINEs~\cite{qel}, although the geometric argument differs: for
$k\geq2$ a semi-algebraic support with empty interior may be a curve, and
the formula $\Lambda$ imposes metric separation instead. The two
satisfiability problems differ in substance: for the two tiles of a
chessboard,
$\Phi_\Gamma$ has an interpretation and no GEL interpretation, since the
alternation along $a+\Real c_1$ cannot be realized by the finite union of
intervals in which a semi-algebraic set meets that line. The semi-algebraicity
of supports does not establish that every
semi-algebraic relation is definable in $\ELR$. The standard translation
makes no claim of surjectivity.

\section{A Vascular Use Case: From Angiographic Annotation to Cohort
Evidence}\label{sec:vascular}

Vascular geometry gives a concrete use for the logic. A medial ball
representation of the lumen supports phenotype queries and supplies evidence
for their verdicts. The formal results of the section are independent of the
dimension; angiographic projections and fundus photographs are the case
$k=2$, tomographic volumes the case $k=3$. The verdict concerns the recorded annotation. Whether a segmentation
represents the anatomy, and whether a geometric criterion answers a clinical
question, require evidence beyond model checking, as for the temporal
phenotypes of QEL~\cite{qel}.

\begin{figure}[htbp]
\centering
\newcommand{\lumen}[1]{\path[#1] plot[smooth cycle,tension=.55] coordinates {
  (-.37,.38) (-.40,.80) (-.33,1.18) (-.22,1.46) (-.13,1.64) (-.06,1.76) (-.13,1.88) (-.22,2.08)
  (-.29,2.38) (-.27,2.72) (-.2,2.93) (0,2.99) (.2,2.93) (.27,2.72) (.29,2.38) (.22,2.08)
  (.13,1.88) (.06,1.76) (.13,1.64) (.22,1.46) (.33,1.18) (.40,.80) (.37,.38) (.27,.1) (0,.02) (-.27,.1)};}
\begin{tikzpicture}[x=1cm,y=1cm,font=\scriptsize,>={Latex[length=1.8mm,width=1.4mm]},
  frame/.style={draw=black!50,line width=.4pt},
  flow/.style={->,line width=.7pt,inkblue},
  box/.style={draw=inkblue,rounded corners=2pt,fill=inkblue!4,align=flush center,inner sep=2.5pt},
  tbox/.style={draw=inkgreen,rounded corners=2pt,fill=inkgreen!5,align=flush center,inner sep=2.5pt}]
\begin{scope}
\clip (0,0) rectangle (2.3,3);
\ifrasters
  \node[inner sep=0pt,anchor=south west] at (-.331,-.762)
    {\includegraphics[height=3.9cm]{\rasterdir 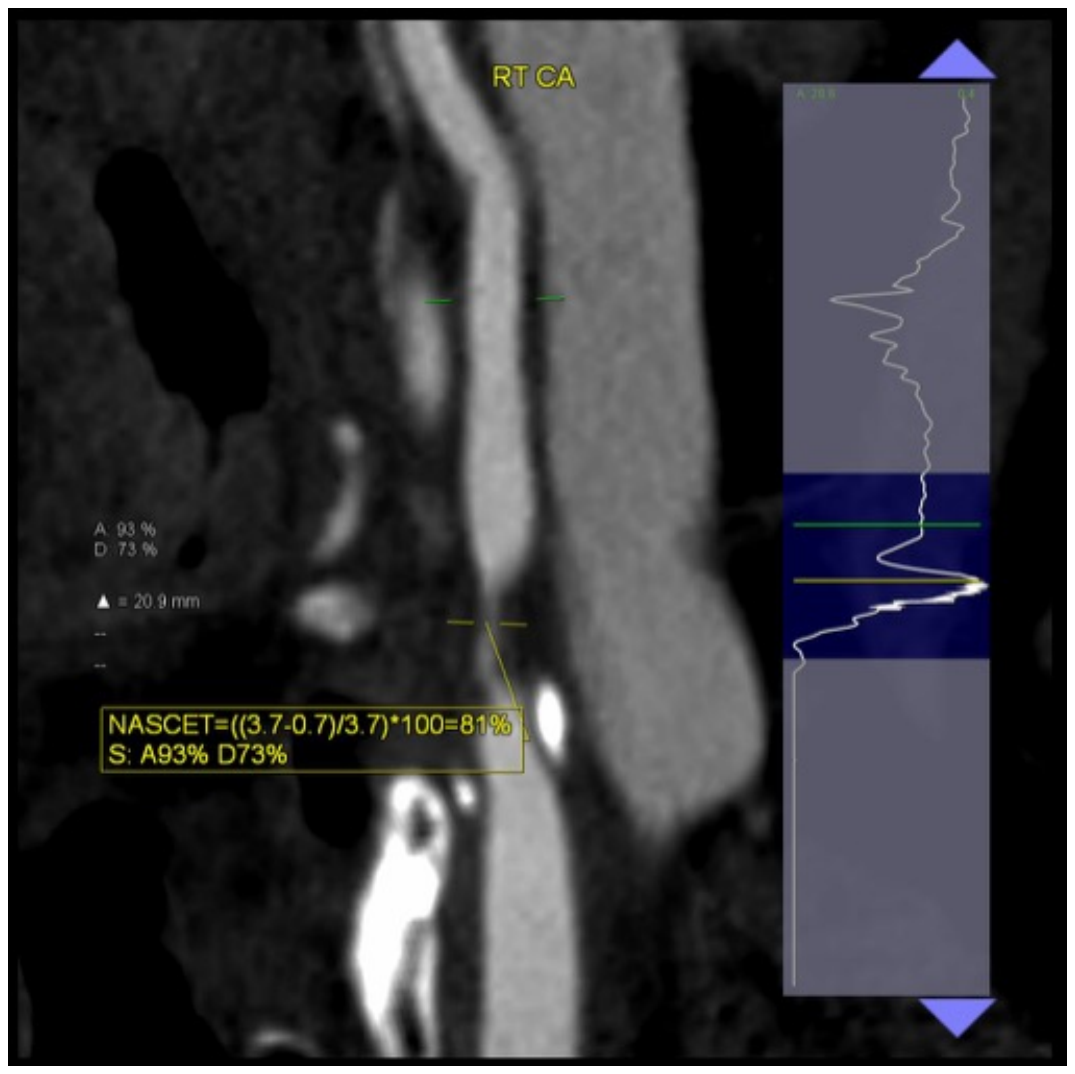}};
\else
  \fill[black!88] (0,0) rectangle (2.3,3);
  \begin{scope}[xshift=1.45cm]\lumen{fill=black!45}\end{scope}
  \draw[yellow!80!orange,line width=.5pt] (1.2,1.76)--(1.7,1.76) (1.05,2.38)--(1.85,2.38);
  \node[draw=yellow!80!orange,text=yellow!80!orange,inner sep=1.3pt,font=\tiny,anchor=west]
    at (.06,1.38) {NASCET $81\%$};
\fi
\end{scope}
\draw[frame] (0,0) rectangle (2.3,3);
\begin{scope}[xshift=2.72cm]
\begin{scope}
\clip (0,0) rectangle (1.3,3);
\fill[black] (0,0) rectangle (1.3,3);
\ifrasters
  \IfFileExists{\rasterdir fig4_vessel3d.pdf}
    {\node[inner sep=0pt,anchor=south west] at (-1.29,-.71)
       {\includegraphics[height=3.903cm]{\rasterdir 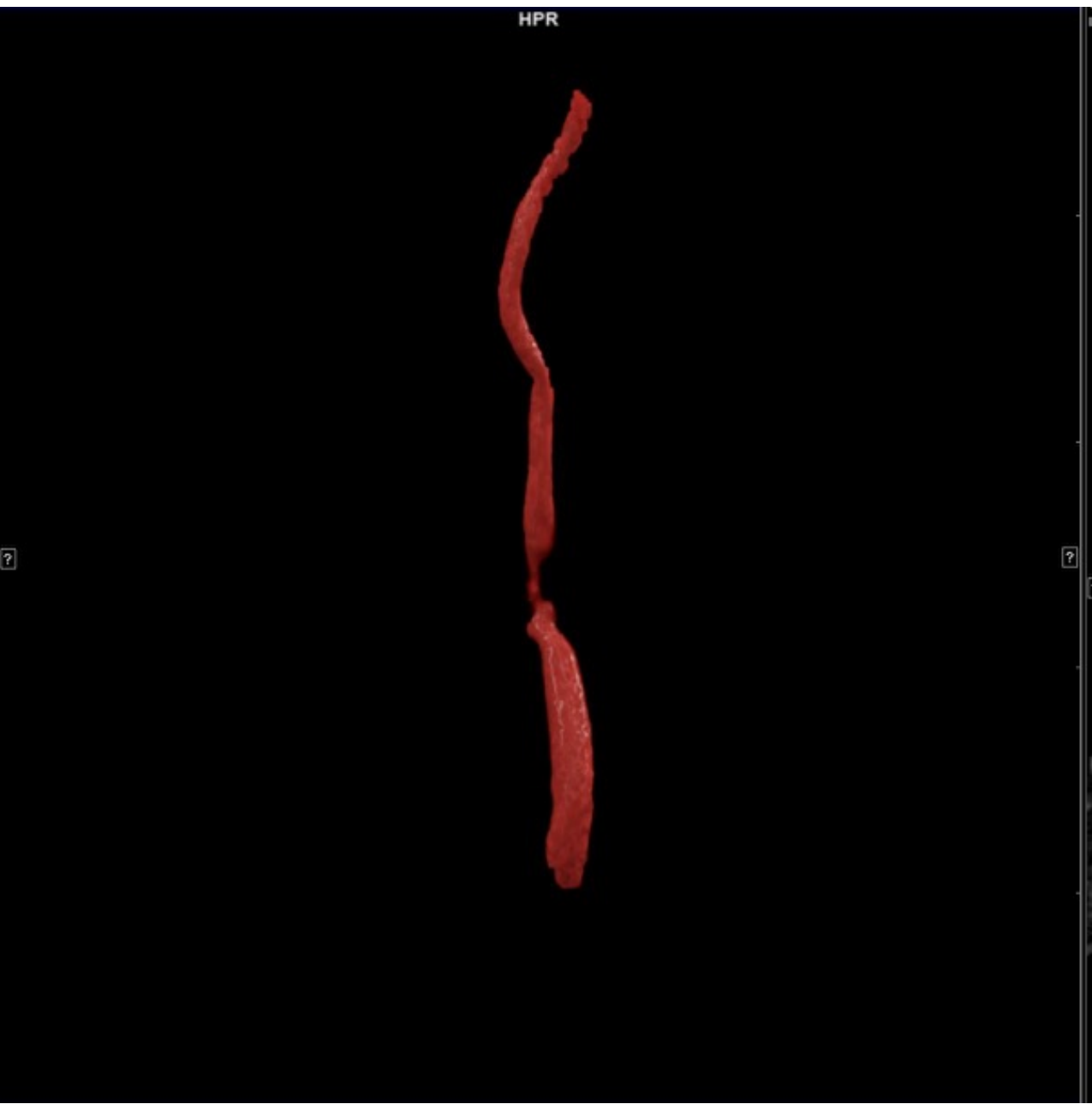}};}
    {\begin{scope}[xshift=.65cm]\lumen{fill=inkred}\end{scope}}
\else
  \begin{scope}[xshift=.65cm]\lumen{fill=inkred}\end{scope}
\fi
\end{scope}
\draw[frame] (0,0) rectangle (1.3,3);
\end{scope}
\begin{scope}[xshift=5.0cm]
\lumen{fill=black!7,draw=black!45,line width=.4pt}
\draw[black!45,line width=.3pt,densely dotted] (0,.1)--(0,2.9);
\foreach \y/\r in {.38/.34,.80/.36,1.18/.30,1.46/.20,1.64/.12,1.88/.12,2.08/.20,2.72/.25}
  {\draw[inkblue,line width=.45pt,fill=inkblue!18,fill opacity=.55] (0,\y) circle [radius=\r];}
\draw[inkred,line width=.6pt,fill=inkred!30] (0,1.76) circle [radius=.055];
\draw[inkgreen,line width=.6pt,fill=inkgreen!25,fill opacity=.6] (0,2.38) circle [radius=.26];
\foreach \y in {.38,.80,1.18,1.46,1.64,1.76,1.88,2.08,2.38,2.72} {\fill (0,\y) circle [radius=.6pt];}
\draw[inkred,line width=.3pt] (.09,1.76)--(.5,1.76);
\node[right=-1pt,text=inkred] at (.5,1.76) {$c_6$};
\draw[inkgreen,line width=.3pt] (.3,2.38)--(.5,2.38);
\node[right=-1pt,text=inkgreen] at (.5,2.38) {$c_9$};
\end{scope}
\node[box,text width=3.4cm,minimum height=3cm,anchor=south west,align=left] (rec) at (6.25,0)
 {$\varepsilon_i(V)=(D_1,\ldots,D_{10})$,\\
  \quad$D_j=D(c_j,\rho_j)$\\[2pt]
  $\varepsilon_i(C)=(D(c_1,0),\ldots)$\\[2pt]
  $\varepsilon_i(C_s)=(D(c_6,0))$\\
  $\varepsilon_i(C_d)=(D(c_9,0))$\\[2pt]
  $\rho_6=\tfrac7{20}$, $\rho_9=\tfrac{37}{20}$\\[2pt]
  unit (mm), frame, metric\\
  matrix, \mbox{provenance}};
\draw[flow] (2.36,1.5)--(2.66,1.5);
\draw[flow] (4.08,1.5)--(4.5,1.5);
\draw[flow] (5.85,1.5)--(6.19,1.5);
\node[tbox,text width=4.5cm] (prot) at (12.5,2.74) {criterion: stenosis above $70\%$};
\node[box,text width=4.5cm] (form) at (12.5,1.96)
 {$\chi_{3/10}=\exists z\,\bigl(\mdiamond(C_s\wedge\neg\boldBox_{3z}V)\wedge\mdiamond C_d$\\
  $\wedge\ \boldBox(C_d\rightarrow\boldBox_{10z}V)\bigr)$};
\node[box,text width=2.1cm] (mc) at (11.3,1.0) {model checking\\ $(\varepsilon_i,0)\models\chi_{3/10}$\,?};
\node[tbox,text width=2.05cm] (coh) at (13.83,1.0) {cohort\\ $\Cohort_{\mathcal{D}}(\chi_{3/10})$};
\node[tbox,text width=4.5cm] (ev) at (12.5,.2)
 {evidence: $c_6$, $c_9$, $\tfrac7{60}<z\leq\tfrac{37}{200}$};
\draw[flow] (prot)--(form);
\draw[flow] (form.south-|mc.north)--(mc.north);
\draw[flow] (rec.east|-mc.west)--(mc.west);
\draw[flow] (mc)--(coh);
\draw[flow] (mc.south)--(mc.south|-ev.north);
\node at (1.15,-.5) {(a) angiogram};
\node at (3.37,-.5) {(b) lumen};
\node at (5.2,-.5) {(c) medial balls};
\node at (8.0,-.5) {(d) record $\varepsilon_i$};
\node at (12.5,-.5) {(e) specification and evaluation};
\end{tikzpicture}
\caption{From a carotid annotation to cohort evidence (image source: CT Dataset (ID: CTA-Athero-Public, 2024)). (a)~\ifrasters A\else
Schematic of a\fi\ CT angiogram with the diameter measurements used to grade
a stenosis. (b)~The segmented lumen. (c)~A medial ball ensemble: ten
inscribed balls with centres on the centreline, including the stenotic
centre $c_6$ and the distal reference centre $c_9$. (d)~The resulting
record. (e)~The criterion is expressed by the formula $\chi_{3/10}$ of
Section~\ref{subsec:library} and evaluated record by record; the verdicts
determine the cohort and retain their evidence
(Example~\ref{ex:nascet}).}

\label{fig:workflow}
\end{figure}

\subsection{Clinical motivation}\label{subsec:motivation}
Stroke is a leading cause of death and long-term disability~\cite{aha2026},
and vascular geometry is central to the imaging decisions that guide its
acute treatment. In the pooled analysis of five randomized trials of
endovascular thrombectomy for proximal anterior-circulation occlusions,
$46.0\%$ of treated patients were functionally independent at 90 days,
compared with $26.5\%$ of controls~\cite{goyal2016}. Later trials extended
treatment to patients selected by imaging up to 24 hours after the time last
known well~\cite{albers2018,nogueira2018}. The relevant findings include the
site of an occlusion, the degree of a stenosis, the configuration of the
circle of Willis, and the collateral routes that sustain tissue beyond an
occlusion~\cite{liebeskind2003}. The benefit of carotid endarterectomy was
established for symptomatic stenosis of 70 to 99 percent, with severity
expressed as a ratio of lumen diameters at prescribed
sites~\cite{nascet1991}. The retina, developmentally a part of the central
nervous system~\cite{london2013}, offers related evidence noninvasively,
through standardized summaries of arteriolar and venular
calibre~\cite{knudtson2003}. A specification can state these calibres, gaps and proximities explicitly.
In practice, their operational meaning is spread across a grading scale, a
reading protocol and code tied to a segmentation pipeline.

\subsection{Strategy}\label{subsec:strategy}
Figure~\ref{fig:workflow} separates four stages of the carotid example.
\emph{Extraction} converts the lumen segmentation into a ball-GEL of
inscribed balls centred on its medial structure
(Section~\ref{subsec:medial}), with further atomic propositions for the
stored centres, anatomical segments, lesions and the observed region. An
inclusion or an enclosure relates this representation to the source geometry (Section~\ref{sec:enclosures}). \emph{Formalization} expresses the
criterion as a closed formula of $\ELR$ (Section~\ref{subsec:library}).
\emph{Model checking} evaluates the formula on each record at the origin
(Lemma~\ref{lem:unbounded}). Theorem~\ref{thm:translation} establishes the
correctness of this step. The calibre tests reduce to comparisons of stored
numbers (Lemma~\ref{lem:medial}). \emph{Reporting} returns the cohort
with the evidence for each verdict: witnessing centres and lengths, or cells
on which a universal condition was verified (Theorem~\ref{thm:evidence}).
An exclusion phenotype uses the complete-window policy $\Full_r$ of
Section~\ref{subsec:observation}, since $\boldBox_r\neg V$ also holds at
points outside the acquired image.

\subsection{Medial ensembles of a lumen}\label{subsec:medial}
Let $X\subsetneq\Real^k$ be a closed set serving as the \emph{reference
geometry} for the lumen proposition $V$. For a mask, it is the support of
$V$ in the cell reconstruction of Section~\ref{subsec:cells}. The radius of
the largest open ball about $c\in X$ contained in $X$ is $\dist(c,X^{c})$,
the clearance of $c$ in $X$. A ball ensemble
$(D(c_1,\rho_1),\ldots,D(c_n,\rho_n))$ is \emph{medial in $X$} when
$c_i\in X$ and $\rho_i=\dist(c_i,X^{c})$ for every $i$; we call $\rho_i$ the
\emph{calibre} at $c_i$. Since $X$ is closed, the ensemble's union lies
in $X$ and generally covers only a proper subset. With these radii,
the geometry that the ensemble omits has no effect on a calibre query at a
stored centre.

\begin{lem}[Calibre at medial centres]\label{lem:medial}
Let $\varepsilon$ be a GEL in which $\varepsilon(V)$ is medial in $X$. For
every centre $c_i$ of $\varepsilon(V)$ and every length $\ell$,
\[
(\varepsilon,c_i)\models\boldBox_\ell V
\quad\Longleftrightarrow\quad\ell\leq\rho_i
\quad\Longleftrightarrow\quad B(c_i,\ell)\subseteq X .
\]
\end{lem}
\begin{proof}
Since $B(c_i,\rho_i)\subseteq\sem{V}_\varepsilon\subseteq X$, the clearances
satisfy
$\rho_i\leq\dist(c_i,\sem{V}_\varepsilon^{c})\leq\dist(c_i,X^{c})=\rho_i$.
Every element is therefore tight, and both equivalences are instances of
Proposition~\ref{prop:morph}(vi), as in
Proposition~\ref{prop:elementwise}(iv).
\end{proof}

\noindent
A single comparison therefore decides a calibre test at a stored centre,
with the same verdict on the record and on the reference geometry. Between
the stored centres the ensemble may certify a smaller clearance, and for a
straight segment the loss can be calculated.

\begin{lem}[Ball chains]\label{lem:chain}
Let $w$ be a vector of norm $1$, let $\rho>0$, $0<s<2\rho$ and $n\geq1$, and
let $V$ hold exactly on $\bigcup_{j=0}^{n}D(c_0+jsw,\rho)$. Put
$h:=(\rho^2-s^2/4)^{1/2}$. Then $\boldBox_\ell V$ holds at the point
$c_0+\tau w$ of the axis whenever $\ell\leq h$ and
$\ell\leq\tau\leq ns-\ell$. For $k\geq2$ the bound is attained: at the
midpoint $c_0+\frac s2w$, the formula $\boldBox_\ell V$ fails for every
$\ell>h$. For $k=1$ the support of $V$ is an interval, and
$\boldBox_\ell V$ holds at $c_0+\tau w$ if and only if
$\ell\leq\rho+\min\{\tau,ns-\tau\}$.
\end{lem}
\begin{proof}
Write a point as $b=c_0+\lambda w+v$ with $\inner{v}{w}=0$, so that
$\norm{b-c_0-jsw}^2=(\lambda-js)^2+\norm{v}^2$. If $0\leq\lambda\leq ns$ and
$\norm{v}\leq h$, choose $j$ with $|\lambda-js|\leq s/2$; the squared
distance is then at most $s^2/4+h^2=\rho^2$, and $V$ holds at $b$. A point of
the window $B(c_0+\tau w,\ell)$ has $|\lambda-\tau|<\ell$ and
$\norm{v}<\ell\leq h$, hence $0<\lambda<ns$, and $V$ holds at it.
Conversely, let $k\geq2$ and $\ell>h$, and choose $v$ with $\inner{v}{w}=0$
and $h<\norm{v}<\ell$. The point $c_0+\frac s2w+v$ lies in the window at the
midpoint, and its squared distance from the $j$-th centre is
$(j-\frac12)^2s^2+\norm{v}^2>s^2/4+h^2=\rho^2$. For $k=1$ the support
consists of the points $c_0+\lambda w$ with $-\rho\leq\lambda\leq ns+\rho$,
because $s<2\rho$, and the clearance of $c_0+\tau w$ in it is
$\rho+\min\{\tau,ns-\tau\}$.
\end{proof}

For a uniformly sampled straight segment in dimension $k\geq2$, the
certified axial clearance thus varies between $\rho$ at the centres and $h$
at the midpoints. For a relative tolerance $\kappa$ between zero and one,
inclusive, a loss of at most $\kappa$, expressed by $h\geq(1-\kappa)\rho$,
is guaranteed exactly when $s\leq2\rho\,(\kappa(2-\kappa))^{1/2}$. A spacing
of half the radius loses $3.2\%$, and a spacing of one radius $13.4\%$.

To construct a medial ensemble from a mask, one measures the clearance to
the reconstructed background. For grid cells with a diagonal metric matrix
of spacings $\sigma_1,\ldots,\sigma_k$, the clearance of the centre of the
foreground cell with index $j$ is the least value of
$\bigl(\sum_{m}(\max\{0,|j_m-i_m|-\tfrac12\}\,\sigma_m)^2\bigr)^{1/2}$ over
the indices $i$ of the background cells, including cells beyond the image
grid. Its square is rational when the squared spacings are, giving rational
presentations of the balls. Local maxima of the clearance supply
candidate centres, a discrete counterpart of the medial
axis~\cite{blum1967,choi1997}. The covering procedure of
Section~\ref{subsec:cover} or the coverage-axis method~\cite{dou2022}
selects among them.

\subsection{Vascular phenotypes as formulas}\label{subsec:library}
Let $V$ denote a lumen presented by a medial ensemble, and let $C$ hold
exactly at its centres, recorded as balls of radius $0$. Further
atomic propositions describe anatomical segments, lesions $L$, territories
$T$ and the observed region $O$, registered in a common frame.

\emph{Calibre.} The formula $\theta_{\mathrm{cal}}(r):=C\wedge\boldBox_rV$
selects the centres of calibre at least $r$; Lemma~\ref{lem:medial} reduces
its evaluation to comparisons. Interpreting twice the calibre as a vessel diameter requires a
cross-sectional convention, recorded in the provenance of $C$.

\emph{Stenosis grading.} Let $C_s$ hold at the centres of a stenotic segment
and $C_d$ at reference centres in the normal vessel distal to it. For
integers $0<m<n$, define
\[
\chi_{m/n}:=\exists z\,\bigl(\mdiamond(C_s\wedge\neg\boldBox_{mz}V)\wedge
\mdiamond C_d\wedge\boldBox(C_d\rightarrow\boldBox_{nz}V)\bigr).
\]
Suppose that both propositions have nonempty supports and that the distal
reference calibre is positive, as the ratio below requires. Write $\rho_s$
and $\rho_d$ for the least recorded calibres of $C_s$ and $C_d$. By
Lemma~\ref{lem:medial}, a length $z$ satisfies the body precisely when
$\rho_s<mz$ and $nz\leq\rho_d$. Such a length exists exactly when
$\rho_s/\rho_d<m/n$, equivalently when the degree of stenosis
$1-\rho_s/\rho_d$ exceeds $1-m/n$. Subject to the diameter convention above,
$\chi_{3/10}$ thus expresses a stenosis greater than $70\%$ in the sense of
the diameter ratio used in the North American Symptomatic Carotid
Endarterectomy Trial~\cite{nascet1991}. The shared length variable supplies a common unit, allowing rational
thresholds to be expressed by additive terms alone.

\begin{exa}[A carotid stenosis]\label{ex:nascet}
Let the residual lumen at the stenosis have the diameter $0.7$\,mm and the
distal reference lumen the diameter $3.7$\,mm, as in the measurement of
Figure~\ref{fig:workflow}(a), so that $\rho_s=7/20$ and $\rho_d=37/20$ in
millimetres. The body of $\chi_{3/10}$ holds exactly for
$7/60<z\leq37/200$. The interval is nonempty, so that the record satisfies
$\chi_{3/10}$; the degree of stenosis is $1-7/37=30/37$, or $81\%$. The
interval of feasible units is the evidence for the verdict
(Corollary~\ref{cor:range}). Its left endpoint is excluded and its right
endpoint is included, because the window is open.
\end{exa}

\emph{Cutoff.} Let $R$ hold on the expected course of a segment, supplied by
an atlas or by the contralateral side. The formula
\[
\chi_{\mathrm{cut}}(r,s):=\mdiamond\bigl(C\wedge\boldBox_rV\wedge
\mdiamond_s(R\wedge\boldBox_r(O\wedge\neg V))\bigr)
\]
holds when a centre of calibre at least $r$ lies at distance less than $s$
from a point on the expected course whose window of length $r$ is fully
observed and free of lumen. The two windows of length $r$ are disjoint,
so the formula is satisfiable only for $s>2r$. This geometric pattern
of an abrupt termination does not by itself distinguish occlusion,
dissection and failure of contrast filling.

\emph{Variants of the circle of Willis.} For the atomic proposition $A$ of a
component with the expected region $R_A$, absence and hypoplasia at the
calibre $r$ are expressed by
\[
\chi_{\mathrm{abs}}(A):=\boldBox(R_A\rightarrow O)\wedge\neg\mdiamond A
\qquad\text{and}\qquad
\chi_{\mathrm{hyp}}(A,r):=\mdiamond A\wedge
\boldBox\bigl((C\wedge A)\rightarrow\neg\boldBox_rV\bigr).
\]
The first formula requires the expected region to be observed and so
excludes a truncated field of view as evidence of absence. The second
requires every recorded centre of the component to have calibre less than
$r$. It presupposes that such centres are recorded; otherwise the universal
condition is vacuously true. Constant formulas cannot
express connectedness (Corollary~\ref{cor:connected}); the connectivity of
a union of balls can be computed geometrically~\cite{edelsbrunner1995} and
recorded as a derived atomic proposition.

\emph{Relations between vessels and lesions.} For a lesion $L$, a vessel $V$
and a territory $T$, three relations recur:
\begin{align*}
\chi_{\mathrm{near}}(r)&:=\mdiamond\bigl(L\wedge\mdiamond_rV\wedge\boldBox_rO\bigr),\\
\chi_{\mathrm{terr}}&:=\mdiamond L\wedge\boldBox(L\rightarrow T),\\
\chi_{\mathrm{clear}}(r)&:=\mdiamond L\wedge
  \boldBox\bigl(L\rightarrow\boldBox_r(O\wedge\neg V)\bigr).
\end{align*}
The first requires a lesion point with a vessel point at distance less than
$r$ and a fully observed window. The second requires a nonempty recorded
lesion contained in the supplied territory. The third requires an observed,
vessel-free window around every lesion point; the conjunct $\mdiamond L$
excludes an empty lesion mask. By Proposition~\ref{prop:morph}(vi), the window condition
of $\chi_{\mathrm{clear}}$ requires distance at least $r$ from the vessel
region, so that tangency at exactly $r$ is admitted, whereas $\mdiamond_rV$
requires distance less than $r$. The existential counterpart
$\chi_{\mathrm{gap}}(r):=\mdiamond\bigl(F\wedge\boldBox_r(O\wedge\neg V)\bigr)$,
with $F$ holding at admissible centres such as a foveal region, describes an
avascular window.

\subsection{Ranges of parameters}\label{subsec:ranges}
Replacing a fixed threshold by a length variable $z$ lets us ask for the
parameter values at which a phenotype holds. The main theorem gives a finite
description of the answer.

\begin{cor}[Feasible lengths]\label{cor:range}
Let $\chi$ be a formula of $\ELR$ whose only free variable is a length
variable $z$, let $\varepsilon$ be a GEL over $\Real^k$, and let
$a\in\Ralg^k$. The set of the lengths $\ell$ such that
$(\varepsilon,a)\models\chi[\ell]$ is a finite union of points and intervals
with endpoints in $\Ralg\cup\{\infty\}$, and such a description is
computable.
\end{cor}
\begin{proof}
By Theorem~\ref{thm:evidence}(i), the set is a semi-algebraic subset of
$\Real$ with a computable presentation. It is therefore a finite union of
points and open intervals whose endpoints are real roots of the polynomials
of the presentation.
\end{proof}

Proposition~\ref{prop:morph}(iv) makes $\chi_{\mathrm{gap}}(r)$ and
$\chi_{\mathrm{clear}}(r)$ antitone in $r$, so that their feasible sets are
initial segments of $\Real_{>0}$. The formula $\chi_{\mathrm{near}}(r)$, with
its conjunct $\boldBox_rO$, is monotone in neither direction: on
$\varepsilon_{\ast}$ of Example~\ref{ex:running} it holds exactly for
$1<r\leq2$. A smaller window reaches no vessel point; a larger one is
incompletely observed.

\subsection{Resources and evaluation protocol}\label{subsec:protocol}
The use case defines a prospective evaluation; no experiment on patient
data is reported here. Several public resources offer complementary
annotations, although none supplies the full set of labels used above: TopCoW, paired
CT and MR angiography with voxel-level labels for 13 components of the
circle of Willis~\cite{topcow2026}; ISLES~2022, multicentre stroke MRI with
expert lesion annotations~\cite{isles2022}; FIVES, fundus photographs with
pixelwise vessel masks~\cite{fives2022}; OCTA-500, OCT and OCTA volumes and
projections with labels for vessels and the foveal avascular
zone~\cite{octa5002024}, which support comparisons of planar and volumetric
records under the conditions of Theorem~\ref{thm:cylinder}; and VESSEL12,
pulmonary CT with sparse public vessel
annotations~\cite{vessel12,vessel12site}. An evaluation should establish
the fidelity of each record to its source and the agreement of a formula
library fixed in advance with independently adjudicated samples.

\section{Discussion: Element-Level Closure and Mathematical
Morphology}\label{sec:discussion}

A geometric record can be designed in two ways. Its representation may contain the regions that queries produce, as
interval ensembles do on the line. Alternatively, the record may consist of
simple elements, leaving the geometry of a query to its evaluation.
Proposition~\ref{prop:closure} rules out the first approach for ball ensembles
in dimension $k\geq2$. We take the second. Evaluation can often remain on
the elements if the union that forms a support is taken last.

\subsection{Boolean closure and the curse of intersection}\label{subsec:curse}
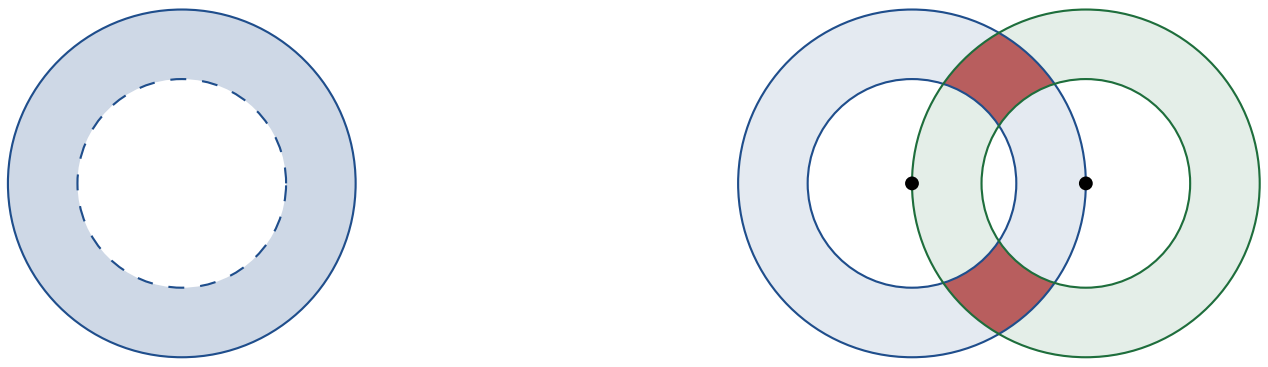
\begin{figure}[htbp]
\centering
\begin{tikzpicture}[font=\small,x=1.15cm,y=1.15cm]
\begin{scope}
\fill[inkblue!22,even odd rule] (0,0) circle (1) (0,0) circle (0.6);
\draw[inkblue] (0,0) circle (1);
\draw[inkblue,dashed] (0,0) circle (0.6);
\node at (0,-1.45){(a) an annulus};
\end{scope}
\begin{scope}[shift={(4.2,0)}]
\fill[inkblue!12,even odd rule] (0,0) circle (1) (0,0) circle (0.6);
\fill[inkgreen!12,even odd rule] (1,0) circle (1) (1,0) circle (0.6);
\begin{scope}[even odd rule]
\clip (0,0) circle (1) (0,0) circle (0.6);
\clip (1,0) circle (1) (1,0) circle (0.6);
\fill[inkred!75] (-1.1,-1.1) rectangle (2.1,1.1);
\end{scope}
\draw[inkblue] (0,0) circle (1) (0,0) circle (0.6);
\draw[inkgreen] (1,0) circle (1) (1,0) circle (0.6);
\fill (0,0) circle (1.3pt) (1,0) circle (1.3pt);
\node at (0.5,-1.45){(b) two annuli};
\end{scope}
\end{tikzpicture}
\caption{Boolean closure applied to disks in the plane. (a)~The disk
$D(0,1)$ without $B(0,3/5)$ is an annulus. (b)~Two such annuli, with centres
at the distance $1$, intersect in two components (dark), each bounded by
four circular arcs.}
\label{fig:intersection}
\end{figure}

Closure under the Boolean operations is the usual requirement on a
representation of regions. Region-based formalisms such as the Region
Connection Calculus~\cite{cohn,aiello2007}, constructive solid
geometry~\cite{requicha1980} and raster masks have it built in, and on the
line interval ensembles possess it (Section~\ref{subsec:closure}). For balls
in dimension $k\geq2$ its price is high. Let $G=I$. The ball $D(0,1)$
without $B(0,3/5)$ is a spherical shell. Two such shells, centred at the
origin and at $e_1$, intersect in two connected components in the plane,
each bounded by four circular arcs (Figure~\ref{fig:intersection}). In a
volume their intersection is a solid ring. Four
balls, intersection and complement have thus produced a cavity, corners
and, in the plane, a disconnected set. This is the \emph{curse of
intersection}: Boolean combinations of simple primitives need not be simple
when constructed explicitly. An intersection of $j$ unions of $n$ balls
distributes into as many as $n^j$ pieces.

\subsection{Deferring the support}\label{subsec:achieves}
A GEL avoids constructing these regions through the division of labour
made explicit by the standard translation. In
Definition~\ref{def:translation} the point of evaluation travels inwards. A
displacement adds its parameter to the evaluation tuple; a bounded modality
replaces the tuple by a point of the window. The record is consulted only at
an atomic proposition, where $\Theta_p$ asks whether the tuple lies in some
element of $\varepsilon(p)$. The union that forms a support is thus taken
at the innermost position. No compound formula's support occurs in
$T_a(\psi)$. General shapes arise when we ask for the points, or the
witnesses, at which a condition holds. A verdict at the point $a$ asks for
less.

Displacements can be brought down to the literals
(Proposition~\ref{prop:laws}(v)). How far the bounded modalities and the
quantifiers can follow them depends on the order in which the operations of
a formula stand between a literal and the root. Call
disjunction, bounded existence and existential quantification the
\emph{existential} operations, and their duals the \emph{universal} ones.
The union over an ensemble is a finite disjunction. It commutes with the
existential operations and displacement, leaving arithmetic on one centre
and one radius
(Proposition~\ref{prop:ball}). A formula built from atomic propositions by
disjunction, displacement and bounded existence therefore holds at a point
exactly when the point lies in a ball with the centre $c-u$ and the radius
$\rho+t$, where $D(c,\rho)$ is an element at some leaf and $u$ and $t$ are
the sums of the displacements and of the lengths above that leaf. The ball
is open if a modality stands above the leaf. This is
Proposition~\ref{prop:elementwise}(i) for closed parameters, and for
arbitrary parameters it holds under each assignment. Dually, a formula built
from negated atomic propositions by conjunction, displacement and bounded
universality holds exactly outside all of these balls. Along such a stretch the element keeps its type. The formula changes the
centre and radius of a ball; the union is taken when the point of evaluation
is tested.

Where the kind of operation changes once, the question still concerns the
elements alone. A conjunction under a bounded existence requires a common
witness: if the supports of $\varphi$ and $\varphi'$ are the unions of balls
$E_1,\ldots,E_n$ and $E'_1,\ldots,E'_{n'}$, then
$\mdiamond_t(\varphi\wedge\varphi')$ holds at $a$ if and only if
$B(a,t)\cap E_i\cap E'_j\neq\emptyset$ for some pair $(i,j)$. This is decided on centres and radii, without constructing or storing the
lens $E_i\cap E'_j$. For a selection of one ball from each of several
conjuncts, Helly's theorem reduces the test to subfamilies of at most $k+1$
balls~\cite{dgk1963}. A disjunction under a bounded universality asks
whether a window is covered by a union of balls. Unless their closures are
pairwise disjoint (Proposition~\ref{prop:elementwise}(iii)), the elements
act jointly here (Example~\ref{ex:coverage}), and the set of points at
which the answer is positive may lie outside $\mathfrak{B}_k$
(Proposition~\ref{prop:closure}). The answer at one point still concerns one
window and finitely many balls.

Quantifiers enter this picture as the modalities do. The points $b$ for
which $p_{nx+u}$ holds at $a$ under $x\mapsto b$ form the balls
$D\bigl((c-a-u)/n,\rho/n\bigr)$, one for each element $D(c,\rho)$ of
$\varepsilon(p)$. An existential point quantifier over a conjunction of
atomic propositions thus asks for a common point. A universal one over a
disjunction of literals asks about a covering. A
quantifier guarded by an atomic proposition whose elements are points ranges
over those points alone. Thus $\mdiamond(C\wedge\varphi)$ holds if and only
if $\varphi$ holds at
some stored centre. A length variable shared by two modalities at one point
is eliminated by a comparison of numbers. By
Proposition~\ref{prop:morph}(vi), the formula
$\exists z\,(\mdiamond_z\varphi\wedge\boldBox_z\varphi')$, with $z$ fresh
for $\varphi$ and $\varphi'$, holds at $a$ exactly when the distance of $a$
from the support of $\varphi$ is less than the clearance of $a$ in the
support of $\varphi'$. On the record $\varepsilon_{\ast}$ of
Example~\ref{ex:running}, $\exists z\,(\mdiamond_zV\wedge\boldBox_zO)$ thus
holds at $a$ if and only if $|a-e_1|<2-|a|$. The support of this formula is
the set of the points $a$ with $|a|+|a-e_1|<2$, for $k\geq2$ the open region
bounded by an ellipsoid with the foci $0$ and $e_1$. Computing the support turns two balls and a shared length into a conic.
Evaluating at a given point needs one comparison. In the stenosis formula of
Section~\ref{subsec:library}, likewise, $C_s$ and $C_d$ confine the point
quantifiers to stored centres, and the length quantifier reduces to the
inequality $\rho_s/\rho_d<m/n$.

The deferral has a limit. When a covering question stands under an
existential operation that ranges over a continuum of points, or a
common-point question under a universal one, the inner verdict becomes a
condition on the outer witness, and the witnesses that satisfy it form sets
of the general kind: $\mdiamond_s\boldBox_1p$, on the ensemble of
Example~\ref{ex:coverage}, asks whether a window meets the cusped set of
Figure~\ref{fig:closure}. For such formulas Theorem~\ref{thm:translation}
remains the general method. It also supplies a correctness criterion for an
evaluator that uses arithmetic for the element-wise stretches, centres and
radii for common points and coverings, and exact arithmetic or certified
enclosures for the remaining constraints.

\subsection{Contrast with mathematical morphology}\label{subsec:contrast}
The asymmetry between dilation and erosion that runs through
Section~\ref{sec:morphology} is the adjunction of
Proposition~\ref{prop:morph}(ii) seen from a particular representation. In
lattice-theoretic morphology, dilation preserves suprema and erosion
preserves infima~\cite{serra1982,heijmansronse1990,heijmans1994}. A ball
ensemble presents its region as the supremum of its elements. Dilation
therefore distributes over the presentation; erosion need not. Morphology
also represents a shape through its skeleton and the radii of its maximal
balls~\cite{blum1967,serra1982,maragos1986}. Exact element-wise erosion
requires of a family $\mathcal{M}$ of closed balls contained in a closed set
$X$ only that every closed ball contained in $X$ lie in a member of
$\mathcal{M}$, as holds for the maximal balls when each such ball lies in a
maximal one. Then
$\boldBox_\ell X$ is the union of the balls $D(c,\rho-\ell)$ over the
members $D(c,\rho)$ of $\mathcal{M}$ with $\rho\geq\ell$, because the
closure of a window $B(a,\ell)\subseteq X$ lies in $X$ and hence in a member
of $\mathcal{M}$, to which Proposition~\ref{prop:ball} applies. In the continuum such a family may be infinite. A medial ensemble is a
finite selection, exact at its stored centres (Lemma~\ref{lem:medial}). On a fixed
raster, masks support the
Boolean operations and digital dilation and erosion without a change of
representation, at a size that depends on the resolution and not on the
depth of the expression evaluated~\cite{soille2004}; the verdicts then
concern the samples (Section~\ref{subsec:cells}).

\subsection{Open problems}\label{subsec:open}
\begin{enumerate}[leftmargin=2em]
\item \emph{Evaluation on the elements.} Is there a calculus for $\ELR$ over
ball-GELs in which every intermediate object is an element or a condition on
a bounded number of elements, as common points and coverings are after one
change of the kind of operation? More modestly, which algorithms decide the
coverage of a window by a union of balls more efficiently than a generic
cylindrical algebraic decomposition, and which fragments admit model
checking in time polynomial in the dimension as well?
\item \emph{Complexity and satisfiability.} What is the exact combined
complexity of model checking, between PSPACE and EXPSPACE
(Proposition~\ref{prop:combined}), and how does it depend on $k$? What is
the exact degree of satisfiability over arbitrary interpretations
(Proposition~\ref{prop:limits}), and is satisfiability decidable for the
closed formulas without quantifiers, whose parameters are constants?
\item \emph{Compression and expressiveness.} Is there a compression theorem
that preserves the truth of a stated class of formulas? How does $\ELR$
compare with the logics of metric spaces~\cite{kutz2003} and with the
spatial logics of closure spaces~\cite{cllm2016}, and is connectedness
definable in the full logic (Corollary~\ref{cor:connected})?
\item \emph{Measurement and space-time.} Volume, surface area and the
length of a centreline involve integration, which lies beyond the
translation of Section~\ref{sec:modelchecking}. Spatiotemporal imaging with
a fixed conversion of units is the present framework in dimension $k+1$;
which results extend to a product of metric monoids whose windows bound
position and time independently, as for the timed words of~\cite{zhang2024temporal}?
\item \emph{Certified evaluation.} Can a proof assistant such as
Lean~\cite{lean4} certify every answer for large collections of records, by
validating real algebraic witnesses and certificates over cells that are
found by an unverified search? A first step has proved feasible: a
formalization in Lean over the library Mathlib~\cite{mathlib} covers the
syntax, the semantics over GELs and the correctness of
Algorithm~\ref{alg:mc} in every dimension and for every metric matrix,
relative to a decision procedure whose verified realization in Lean is the
subject of a public evaluation problem~\cite{leaneval}. Its results,
including the validation of all proofs by the Lean kernel, are reserved for
a separate paper.
\end{enumerate}

\section{Conclusion}\label{sec:conclusion}
Spatial Ensemble Logic gives geometric records a language in which the
meaning of a query is fixed before a method of evaluation is chosen. Its
syntax and semantics are those of the member $\ELR$ of $\EL$ over Euclidean
space $\Real^k$ with open-ball windows, and geometric element ensembles are
the finite presentations of its interpretations. The standard translation
into real closed fields connects the two. It yields decidable model
checking, finite real algebraic evidence and, under the stated fixed-query
and input assumptions, polynomial data complexity. The same connection
yields synthesis within a prescribed budget of balls. Removing the budget leads to undecidability. Enclosures allow approximate records to
support sound logical verdicts.

The geometric representation and the domain of evaluation need not
coincide. On the line they do: finite unions of intervals remain closed
under the relevant operations, windows admit finite subdivision, and
rational interval input admits linear evaluation and rational evidence under
the rationality assumptions of Proposition~\ref{prop:linear}. For $k\geq2$,
Euclidean balls no longer provide those general guarantees of closure and
rationality, yet individual balls keep a simple calculus of translation,
dilation and erosion in every dimension. A record can therefore consist of simple elements while algebraic
evaluation handles the Boolean and quantified conditions of a query.
Deferring the union to the point of evaluation lets much of the computation
remain on the elements themselves, as in the calibre and stenosis queries
of the vascular use case.

\section*{Acknowledgments}
The author thanks Licong Cui, Manfred Droste, Baowei Fei, Xubing Hao, Yan Huang, Xiaojin Li, and Sunil Sheth
for feedback and illuminating discussions. This work has been supported in part by the U.S.
National Science Foundation Award IIS-2500624 and the National Institutes of Health grants R01AG084236 and
U24AG098157. The views expressed in this paper are those
of the author and do not necessarily reflect those of the funding agencies.

\end{document}